\PassOptionsToPackage{table}{xcolor}
\documentclass[fleqn]{article}
\immediate\write18{texcount -inc -sub=subsection test.tex > wordcount.txt}
\usepackage{graphicx} 
\usepackage{subfig}
\usepackage{authblk}
\usepackage{amsmath}
\usepackage[dvipsnames]{xcolor}
\usepackage{hyperref}
\usepackage[margin=1.2in]{geometry}
\usepackage{amssymb}
\usepackage{booktabs} 
\usepackage{subcaption}
\usepackage{cite}
\usepackage{lineno}

\usepackage{booktabs}
\usepackage[table]{xcolor}
\usepackage{array}
\usepackage{makecell}
\usepackage{siunitx}

\usepackage{xcolor}

\definecolor{NatureHeader}{RGB}{225,228,214}
\definecolor{NatureStripe}{RGB}{245,246,240}

\usepackage{xcolor}
\usepackage{booktabs}
\usepackage{makecell}
\usepackage{threeparttable}

\definecolor{NatureHeader}{RGB}{225,228,214}

\title{Pathway-resolved analysis of internal conversion enabled by Gaussian boson sampling}

\author[1,2]{Mikhail Roiz\thanks{Corresponding author, mikhail.roiz@uni-paderborn.de}}
\author[3]{Rashid Valiev\thanks{Corresponding author, rashid.valiev@helsinki.fi}}
\author[1,2]{Jan-Lucas Eickmann}
\author[3]{Theo Kurten}
\author[1,2]{Benjamin Brecht}
\author[1,2]{Michael Stefszky}
\author[1,2]{Christine Silberhorn}
\affil[1]{Paderborn University, Integrated Quantum Optics, Institute for Photonic Quantum Systems (PhoQS), Warburger Str. 100, 33098 Paderborn, Germany}
\affil[2]{Paderborn University, Institute for Photonic Quantum Systems (PhoQS), Warburger Str. 100, 33098, Paderborn, Germany}
\affil[3]{Department of Chemistry, Faculty of Science, P.O. Box 55 (A.I. Virtanens plats 1),
FIN-00014 University of Helsinki, Finland}

\begin{document}
\maketitle

\begin{abstract}
Gaussian Boson Sampling (GBS) has provided strong evidence of quantum advantage, motivating the search for real-world applications. Such applications rely on mapping physical problems to the native GBS output: photon-number patterns sampled from a highly non-trivial probability distribution. Here we derive a GBS-compatible mapping for the rate constant $k_\mathrm{IC}$ of internal conversion (IC), incorporating two important physical effects -- Duschinsky rotations and Herzberg--Teller contribution. We validate this framework using sampling \textit{simulations} of dibenzoterrylene, reproducing reported values of $k_\mathrm{IC}$. Beyond $k_\mathrm{IC}$ reconstruction, we show that the photon-number patterns of GBS encode the vibrational mode combinations responsible for the IC process, which we refer to as non-radiative pathways. This pathway-resolved data can then be grouped in different ways to extract multiple physically motivated observables. In particular, it enables $k_\mathrm{IC}$ analysis at several levels, from individual modes and mode family contributions to recurring vibrational configurations shared by multiple pathways. We find that Duschinsky rotations increase the fraction of $k_\mathrm{IC}$ associated with correlated vibrational pathways from 46\% to 86\%, and identify the mode combinations with the largest rate contributions. These pathway-resolved insights can guide targeted molecular modifications, including selective deuteration and bond substitution, to control non-radiative decay.
\end{abstract}

Gaussian Boson Sampling (GBS) \cite{Hamilton2017,KruseGBS} has been established as a scalable, non-universal photonic quantum computing platform for demonstrating quantum advantage --  a scenario in which a quantum computing device outperforms all possible classical algorithms for a given task \cite{Zhong2020, Liu2026Jiuzhang4,Zhong2021,Madsen2022,Deng2023,Yu2023,Zhu2024}. An important next step is therefore to extend existing demonstrations of quantum advantage towards practically relevant tasks. This has motivated an active search for applications that naturally map onto the GBS platform \cite{huh2015boson,Huh2017VibronicGBS,Arrazola2021,Bradler2021,Deng2023Graph,Anteneh2023,Nokkala2021,Hoch2025,Cimini2025,Sakurai2025}.

Such mappings rely on the native output of GBS, which is the sampling of photon-number patterns from a highly non-trivial probability distribution \cite{Hamilton2017,KruseGBS}. One of the first applications that aimed to exploit this property was the simulation of molecular vibronic spectra  \cite{huh2015boson,Huh2017VibronicGBS}. In this application, each photon-number pattern can be mapped to a specific vibronic transition, described by a distinct combination of vibrational quanta among molecular modes. The strength of each specific transition is given by a Franck--Condon factor (FCF), which under the GBS mapping can be estimated from the probability of the corresponding photon-number pattern \cite{huh2015boson,Huh2017VibronicGBS}. The complete set of these transitions defines a multi-dimensional observable -- the \textit{mode-resolved vibronic spectrum} -- which contains the exact combinations of vibrational quanta among the modes.

Reconstructing the full mode-resolved spectrum with GBS, however, requires exponentially many samples as molecular size grows \cite{Oh2021, Seron2026}. This necessitates \textit{grouping} sampled photon patterns into lower-dimensional quantities defined by the target physical observable \cite{Oh2021,Seron2026}. In spectroscopy, this observable is usually the spectral envelope (or Franck-Condon profile) \cite{huh2015boson,Oh2021}, obtained by grouping individual vibronic transitions into finite-width energy bins, effectively erasing information about combinations of vibrational quanta among modes \cite{Oh2021}. While this grouping overcomes the issue of needing exponentially many samples, the reduced task no longer exploits the complete information contained in photon-number patterns -- in fact, it can be calculated efficiently using classical Fourier-transform methods \cite{Oh2021,Seron2026,Oh2024,Oh2024Alg,Lim2025}. This raises the question of whether the information encoded in photon-number patterns can be exploited further using different groupings.

Here we move beyond the molecular vibronic spectra setting by applying alternative groupings of GBS data for pathway-resolved analysis of the internal-conversion (IC) rate constant $k_\mathrm{IC}$. IC is one of the two principal non-radiative relaxation processes in molecules \cite{Valiev2018}, alongside intersystem crossing \cite{Valiev2018,Valiev2020,Khoroshkin2026OzoneGBS}. During IC, the initial electronic excitation energy is redistributed among combinations of molecular vibrational modes without emitting radiation. We refer to these combinations as non-radiative pathways, represented by distinct output photon-number patterns in GBS.

For applications where light emission is of primary interest, molecules with high fluorescence quantum yields are generally desirable \cite{ValievDBT,Chen2026Deuteration,Tuckova2022INVEST,Toews2026}. Since IC competes with fluorescence, resolving the vibrational pathways contributing to $k_\mathrm{IC}$ can identify targeted molecular modifications for suppressing this loss channel \cite{ValievDBT,Chen2026Deuteration,Tuckova2022INVEST,Toews2026}. Such analysis requires a balanced treatment of the three physical effects -- Duschinsky rotations, Herzberg–Teller contributions, and anharmonicity \cite{Valiev2018}. Obtaining the complete pathway information becomes challenging specifically when the Duschinsky mode-mixing is included, since it prevents a simple decomposition into independent-mode contributions. This challenge makes the pathway-resolved IC problem a natural target application for the GBS architecture.

In this work, we derive an explicit mapping of the IC problem onto a GBS-compatible sampling procedure that directly incorporates two important physical effects -- Duschinsky rotations and HT contribution. We validate the developed GBS framework by performing sampling \textit{simulations} of dibenzoterrylene (DBT) -- a molecule with unique emission properties \cite{JelezkoDBT1997, WangDBT2017,WangDBT2019,PschererDBT2021, MishraDBT2025} and a large number of vibrational modes (168). We show that our method not only reproduces $k_\mathrm{IC}$ values obtained in the previous studies \cite{ValievDBT,Toninelli2021}, but also provides new and unique information about the IC process itself. The central insight of our work is that the photon-number patterns produced via sampling correspond directly to the pathway-resolved information of the non-radiative decay and that this information can be used to extract multiple physically motivated observables. In particular, we show that different groupings of the sampling data enable $k_\mathrm{IC}$ analysis at several levels, from the contributions of individual vibrational modes and mode families to the recurring vibrational configurations shared by multiple pathways. Our analysis reveals that in the Duschinsky model over 86\% of $k_\mathrm{IC}$ is associated with pathways containing correlated mode combinations, predominantly of second order, compared with only 46\% in the absence of Duschinsky rotations. These results highlight that for rational strategies of $k_\mathrm{IC}$ suppression it is important to consider correlated vibrational structures rather than treating the modes independently. To our knowledge, this is the first approach that enables the full pathway-resolved analysis of $k_\mathrm{IC}$ in a computationally demanding regime where both HT and Duschinsky mode-mixing contributions are treated simultaneously and explicitly.

\subsection*{Mapping $k_\mathrm{IC}$ onto Gaussian boson sampling}

IC describes the non-radiative intramolecular relaxation process and is mainly characterized by its rate constant $k_{\mathrm{IC}}$. In this process, the energy contained in the pre-excited molecule is converted into the molecular vibrations, which are described by phonons -- bosonic quasi-particles that represents quantized lattice vibrations. Since photons and phonons are bosons, this enables mapping of the vibrational relaxation processes onto the photonic sampling platforms such as GBS.

The rate constant $k_\mathrm{IC}$ can be calculated using Fermi's golden rule, however, for realistic polyatomic molecules it is commonly described by a Lorentzian-broadened expression first introduced by Plotnikov \cite{Plotnikov1979}
\begin{equation}\label{k_Plotnikov}
k_{\mathrm{IC}}
=
\sum_{n,m^\prime}\left|\langle \psi_{f}, \boldsymbol{m^\prime} \lvert \hat{V} \rvert \psi_{i} ,\boldsymbol{n}\rangle\right|^2
\left[
\frac{\Gamma_{fm^\prime}}{\Delta_{fi}^2+\Gamma_{fm^\prime}^2/4}
\right].
\end{equation}
Here $|\psi_{i}\rangle$ and $|\psi_{f}\rangle$ are the initial and final electronic eigenstates, $|\boldsymbol{n}\rangle = | n_{1}, \dots, n_{k}, \dots, n_{M} \rangle $ and $|\boldsymbol{m^\prime}\rangle = | m_{1}^\prime, \dots, m_{k}^\prime, \dots, m_{M}^\prime\rangle $ are the initial and final vibrational eigenstates and the prime sign indicates the normal-mode basis of the final electronic state. Index \textit{M}=3\textit{N}-6 (\textit{M}=3\textit{N}-5 for linear molecules) is the number of normal modes with $N$ being the number of atoms in the molecule \cite{Valiev2020, EickmannVibronic2025}. The operator $\hat{V}$ is the non-adiabatic coupling (NAC) operator and the term $\langle \psi_{f}, \boldsymbol{m^\prime} \lvert \hat{V} \rvert \psi_{i}, \boldsymbol{n}  \rangle = V_{i\boldsymbol{n},f\boldsymbol{m}^\prime}$ represents the corresponding non-adiabatic coupling matrix element (NACME), whose indices specify the non-radiative pathway and its magnitude determines the coupling strength. $\Delta_{fi}=\lvert E_{fm^\prime}-E_{in}\rvert$ is the energy difference between the initial and final electronic states. Each vibrational mode $k$ is characterized by the corresponding number of vibrational quanta $n_k$ and $m_k^\prime$, which are represented by the number of photons in each output mode of the GBS device. In this work we will consider the zero-temperature approximation, meaning that the initial vibrational quantum number is $|\boldsymbol{n}\rangle = |\boldsymbol{0}\rangle$ as no phonons are excited in the initial state \cite{ValievDBT}. $\Gamma_{fm^\prime}$ is a phenomenological broadening parameter that replaces the Dirac delta function in Fermi's golden rule by a Lorentzian lineshape of width $\Gamma_{fm^\prime}$. Assuming that $\Gamma_{fm^\prime}$ does not depend on the final vibrational state, equation \eqref{k_Plotnikov} is commonly simplified using the flat-window approximation as \cite{Valiev2018,Valiev2020}
\begin{equation}\label{eq:k_flat_window}
k_{\mathrm{IC}}
=
\frac{4}{\Gamma_{fm^\prime}}\sum_{\boldsymbol{m}^\prime\in Q_{\Delta E}}\left|\langle \psi_{f}, \boldsymbol{m^\prime} \lvert \hat{V} \rvert \psi_{i} ,\boldsymbol{0}\rangle\right|^2.
\end{equation}
This formula treats the set of transitions $Q_{\Delta E}$ landed in the energy window $\pm \Delta E$ equally without applying Lorentzian lineshape, while transitions outside the window are discarded \cite{ValievDBT}.

To make the description of non-radiative transitions more convenient, one can define a set of mass-weighted normal coordinates \cite{duschinsky1937interpretation} $\boldsymbol{q} = (q_{1}, \dots, q_{M}) $ and $\boldsymbol{q^\prime} = (q_{1}^\prime, \dots, q_{M}^\prime) $ for the initial and final states, respectively. The relationship between these coordinates is given by the Duschinsky transformation $\boldsymbol{q}^\prime = \mathbf{U}\boldsymbol{q} + \Delta \boldsymbol{q}$ \cite{duschinsky1937interpretation}, where $\Delta\boldsymbol{q} =  (\Delta q_{1}, \dots, \Delta q_{M})$ is the nuclei displacement vector that represents the shift along the normal coordinates and the Duschinsky matrix $\mathbf{U}$ describes mixing between the normal coordinates \cite{duschinsky1937interpretation}. 

NACME can also be expressed using the Doktorov operator as $V_{i\boldsymbol{0},f\boldsymbol{m}}=\langle \psi_{f}, \boldsymbol{m} \lvert \hat{U}_{Dok}  \hat{V} \rvert \psi_{i}, \boldsymbol{0}\rangle$,  which transforms the final vibrational state to the same basis as the initial vibrational state $|\boldsymbol{m^\prime}\rangle = \hat{U}_{Dok}^\dagger |\boldsymbol{m}\rangle$ \cite{DOKTOROV}. A convenient way to describe this transformation is to decompose it into the building blocks of GBS as $\hat{U}_{Dok} = \hat{D}(\boldsymbol{\beta}) \hat{R}(\mathbf{U_L})\hat{S}(\boldsymbol{r})\hat{R}(\mathbf{U_R})$, with $\hat{D}$, $\hat{S}$ and $\hat{R}$ being the displacement, squeezing and rotation operators \cite{huh2015boson}. The displacement operator parameter $\boldsymbol{\beta}$ is connected to the nuclei displacement as $\beta_k =\sqrt{\frac{\omega^\prime_k}{2\hbar}} \Delta q_k$ \cite{EickmannVibronic2025}, whereas other operator parameters are linked to the matrix $\mathbf{J} = \Omega^{\prime}\textbf{U}\Omega^{-1}$, where \textbf{U} is the Duschinsky matrix, $\Omega = \text{diag}(\sqrt{\omega_1}, ..., \sqrt{\omega_M})$ and $\Omega^{\prime} = \text{diag}(\sqrt{\omega_1^{\prime}}, ..., \sqrt{\omega_M^{\prime}})$ are the diagonal matrices of initial and final vibrational frequencies, respectively \cite{EickmannVibronic2025}.
The direct connection between these parameters and $\hat{U}_{Dok}$ can be established using the singular value decomposition $\mathbf{J}=\Omega^{\prime}\textbf{U}\Omega^{-1} = \mathbf{U_L}\mathbf{\Sigma}\mathbf{U_R}$ \cite{huh2015boson,EickmannVibronic2025}, where the diagonal matrix of singular values $\mathbf{\Sigma}$ defines the squeezing parameters as $r_k = \ln{\left(\Sigma_{k,k}\right)}$, while the two rotation operators with the corresponding unitaries $\mathbf{U_L}$ and $\mathbf{U_R}$ account for the mixing of optical modes. It is important to note that, within the GBS framework, the Duschinsky rotations are represented through the combined action of the squeezing and rotation operators \cite{EickmannVibronic2025}.

Thus far, we provided mapping only to the Doktorov transformation, which describes the conversion between bases of the initial and final states. To complete the mapping, however, one must also consider the explicit form of the NAC operator $\hat V$. This operator determines the coupling between the two electronic states and can be written in the following form \cite{DomckeStock1997}
\begin{equation} \label{eq:Dcoupling1}
\hat V=\sum_{i\neq f}(|\psi_{f}\rangle\,\hat\Lambda_{fi}\,\langle\psi_{i}|
+\mathrm{h.c.}),
\qquad
\hat\Lambda_{fi}=-\hbar^2\sum_{k=1}^M
g_{fi,k}(\mathbf q)\,\frac{\partial}{\partial q_k}
-\frac{\hbar^2}{2}\sum_{k=1}^M
h_{fi,k}(\mathbf q),
\end{equation}
where $\hat\Lambda_{fi}$ describes the nuclear coordinate dependence of the NAC operator. Here, index $k=1,\ldots,M$ represents the normal mode number, with $M$ being the total number of modes. The first and second order derivative coupling components, also known as electronic matrix elements, are expressed as
\begin{equation} \label{eq:DcouplingExpl}
g_{fi,k}(\mathbf q)
=
\left\langle \psi_{f}(\mathbf q)\left|\frac{\partial}{\partial q_k}\right|\psi_{i}(\mathbf q)\right\rangle,
\qquad
h_{fi,k}(\mathbf q)
=
\left\langle \psi_{f}(\mathbf q)\left|\frac{\partial^2}{\partial q_k^2}\right|\psi_{i}(\mathbf q)\right\rangle.
\end{equation}
The explicit evaluation of these matrix elements requires suitable approximations. As a first step, we neglect the $h_{fi,k}(\mathbf q)$ term, whose contribution has been shown to be negligible \cite{Valiev2018,Valiev2020}. The first derivative coupling term can be Taylor-expanded around the reference molecular geometry $\mathbf q_0$ as
\begin{equation}\label{eq:g_HT_expand}
g_{fi,k}(\mathbf q)
\approx
g_{fi,k}(\mathbf q_0)+\sum_{j=1}^M \left.\frac{\partial g_{fi,k}(\mathbf q)}{\partial q_j}\right|_{\mathbf q_0}\,\Delta q_j,
\end{equation}
where $\Delta q_j=q_j-q_{0,j}$. Assuming zero temperature $|\mathbf n\rangle=|\mathbf 0\rangle$ and introducing the Doktorov operator, the corresponding NACME becomes
\begin{equation}\label{eq:NACME_HT_Dok}
V_{i \boldsymbol 0,f\boldsymbol m}
=
\sum_{k=1}^M \Tilde{y}_k\;
\langle \boldsymbol{m} |\hat U_{{Dok}}(\hat a_k^\dagger-\hat a_k)|\mathbf 0\rangle
+
\sum_{j=1}^M\sum_{k=1}^M \Tilde{z}_{j,k}\;
\langle \boldsymbol{m} |\hat U_{{Dok}}(\hat a_j^\dagger+\hat a_j)(\hat a_k^\dagger-\hat a_k)|\mathbf 0\rangle,
\end{equation}
where the coordinate derivative and displacement were expressed in terms of the ladder operators as $\frac{\partial}{\partial q_k}=-\sqrt{\frac{\omega^\prime_k}{2\hbar}} (\hat a_k^\dagger- \hat a_k)$, $\Delta q_j = \sqrt{\frac{\hbar}{2\omega^\prime_j}}(\hat a_j^\dagger+\hat a_j)$. The parameters $\Tilde{y}_k \in \mathbb{R}$ and $\Tilde{z}_{jk} \in \mathbb{C}$ carry the actual physical dimensions and are defined as $\Tilde{y}_k = \hbar^2\sqrt{\frac{\omega^\prime_k}{2\hbar}} g_{fi,k}(\mathbf q_0)$, $\Tilde{z}_{j,k}=\frac{\hbar^2}{2}\left.\sqrt{\frac{\omega^\prime_k}{\omega^\prime_j}}\frac{\partial g_{fi,k}(\mathbf q)}{\partial q_j}\right|_{\mathbf q_0}$.

The first term in equation \eqref{eq:NACME_HT_Dok} corresponds to the Condon approximation, whereas the second term is associated with the HT contribution \cite{Valiev2018}. Each term involves the direct action of $(\hat a_k^\dagger-\hat a_k)$ or $(\hat a_k^\dagger+\hat a_k)$, neither of which is a unitary Gaussian transformation, making them incompatible with the GBS platform. One way to solve this problem is to approximate the nonunitary part as a linear superposition of unitary Gaussian operations, which was suggested by Jnane \textit{et al.} \cite{Jnane2021}. In this approach, one needs to introduce an auxiliary function
\begin{equation} \label{eq:Aux_f}
f_{m}(\xi)
=
|\langle \boldsymbol{m} |\hat{U}_{Dok} e^{\xi \hat{G}} |\mathbf 0\rangle|^2,
\end{equation}
with the expansion coefficient $\xi \in \mathbb{C}$ \cite{Jnane2021}. Here, the auxiliary operator $\hat{G}$ represents the corresponding nonunitary operation
\begin{equation}\label{eq:G_aux}
\hat G
=
\sum_{k=1}^M
y_k(\hat a_k^\dagger-\hat a_k)
+
\sum_{j=1}^M\sum_{k=1}^M z_{j,k}
(\hat a_j^\dagger+\hat a_j)(\hat a_k^\dagger-\hat a_k),
\end{equation}
where we define $y_k = \Tilde{y}_k/p$ and $z_{j,k} = \Tilde{z}_{j,k}/p$ with $p$ being the normalization factor with the same dimensions as $\Tilde{y}_k$ and $\Tilde{z}_{j,k}$ to make the quantities dimensionless. Expanding \eqref{eq:Aux_f} with respect to $\xi$ allows us to approximate the square of NACME values as a superposition of the auxiliary functions
\begin{equation} \label{eq:Superposition}
|\langle \boldsymbol{m} | \hat{U}_{Dok}\,\hat{G} | \boldsymbol{0} \rangle|^{2}
\approx
\frac{1}{2\tau^{2}}
\left[
f_{\boldsymbol{m}}(i\tau)+\frac{1}{2}f_{\boldsymbol{m}}(\tau)+\frac{1}{2}f_{\boldsymbol{m}}(-\tau)-2f_{\boldsymbol{m}}(0)
\right] + O(\tau^2),
\end{equation}
where $\tau$ is a real positive number defining the expansion coefficient values $\xi=\{i\tau, \tau, -\tau, 0\}$ \cite{Jnane2021}.

It is convenient to write \eqref{eq:G_aux} in the vector-matrix form
\begin{equation}\label{eq:Ggen_canonical_full}
\hat G
=
\mathrm{Tr}(\mathbf{z})
+
\left(\boldsymbol{y}^T\hat{\mathbf a}^\dagger-\boldsymbol{y}^T\hat{\mathbf a}\right)
+
\frac{1}{2}\left(
\hat{\mathbf a}^{\dagger T}\mathbf{z}_s\hat{\mathbf a}^\dagger
-\hat{\mathbf a}^{T}\mathbf{z}_s\hat{\mathbf a}\right)
-\hat{\mathbf a}^{\dagger T}\mathbf{z}_a\hat{\mathbf a}
=
\mathrm{Tr}(\mathbf{z})+ \hat G_0,
\end{equation}
where we introduced multi-mode ladder operators $\hat{\mathbf{a}}=(a_1, \dots, a_M)^T$, $\hat{\mathbf a}^\dagger=(a_1^\dagger, \dots, a_M^\dagger)^T$, and the dimensionless vector $\boldsymbol{y}=(y_1, \dots, y_M)^T$. The dimensionless matrix $\mathbf{z}=(z_{j,k})$ is decomposed into its symmetric $\mathbf{z}_s=\mathbf{z}+\mathbf{z}^T$ and anti-symmetric $\mathbf{z}_a=\mathbf{z}-\mathbf{z}^T$ counterparts. From equation \eqref{eq:Ggen_canonical_full} it is clear that the operator $\hat G_0$ has at most quadratic dependence upon the ladder operators. This means that $\hat G_0$ represents a Gaussian transformation, therefore it can be mapped onto the GBS platform. This mapping could potentially be obtained by a modified version of the method described in \cite{Jnane2021}, but instead here we adapt the approach of nullifier equations \cite{FunaiNULLIFIERS2025,LasserNULLIFIERS2018}. 

We consider the state $|\psi_\xi^{(0)}\rangle = e^{\xi \hat{G_0}} |\mathbf 0\rangle$, which can be decomposed into the building blocks of GBS as
\begin{equation}\label{eq:aux_state}
|\psi_\xi^{(0)}\rangle
=
C_\xi\,
\hat D(\bar{\boldsymbol\gamma}_\xi)\,
\hat R(\bar{\mathbf U}_\xi)\,
\hat S(\bar{\mathbf\Sigma}_\xi)\,
|\mathbf 0\rangle,
\end{equation}
where the right most rotation operator has been dropped, since it acts trivially on vacuum. The operator parameters entering equation \eqref{eq:aux_state} can be found by considering the corresponding nullifier equation and the subsequent Takagi decomposition \cite{CariolaroTAKAGI}. The detailed derivation of the auxiliary operator parameters and the prefactor $C_\xi$ can be found in the Methods section and Supplementary Information.

\begin{figure}[!ht]
    \centering
    \includegraphics[width = 1\textwidth]{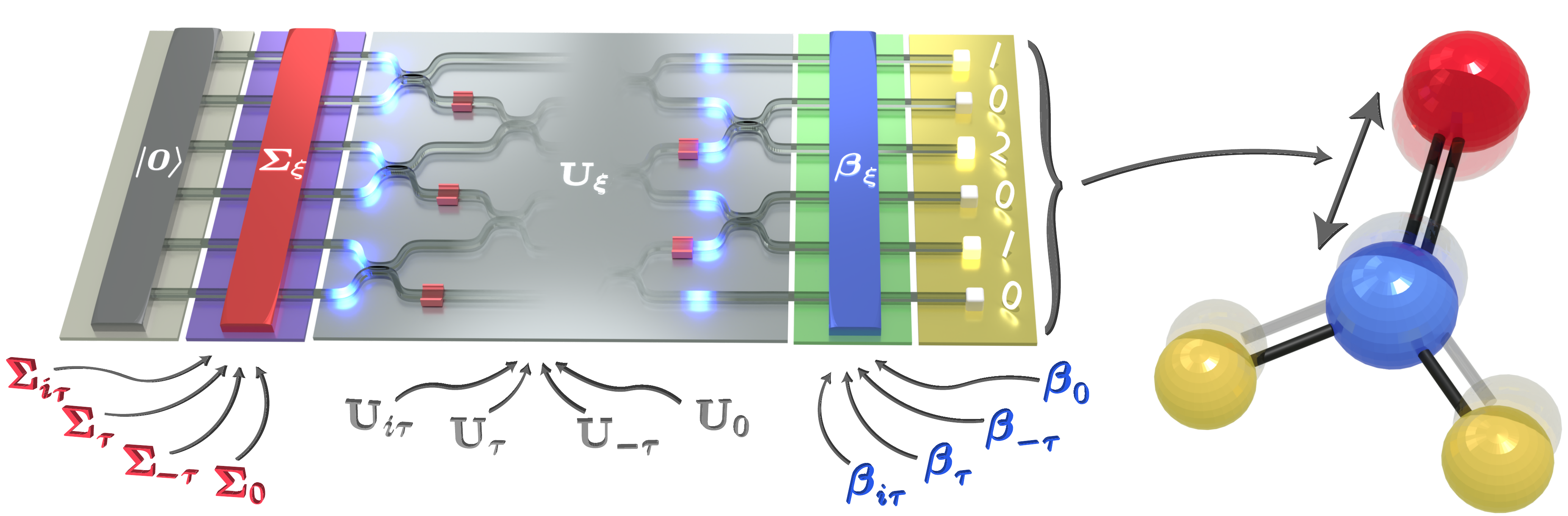}
    \caption{General schematic of the Gaussian boson sampling (GBS) based framework for $k_\mathrm{IC}$. The GBS experiment is performed four times with four different sets of input optical parameters for the squeezing $\hat S(\boldsymbol\Sigma_\xi)$, rotation $\hat R(\mathbf U_\xi)$ and displacement $\hat D(\boldsymbol\beta_\xi)$ operators. These sets correspond to the expansion coefficient values $\xi=\{i\tau, \tau, -\tau, 0\}$. At the final stage of the GBS experiment, the photon-number-resolving detection is employed to reconstruct the photon pattern distributions. The statistics of the four sampling experiments are used to evaluate the non-adiabatic coupling matrix elements $|V_{i \boldsymbol 0,f\boldsymbol m}|^2$ according to equation \eqref{eq:Superposition}. As schematically shown in the figure, each photon pattern generated in GBS corresponds to the excitation of a distinct set of vibrational modes in a molecule. This photon-pattern-resolved information is used to perform the vibrational mode-resolved analysis of the non-radiative pathways in the molecule.}
    \label{Fig:GBS_concept}
\end{figure}

The full form of the auxiliary function can be obtained by expanding the Doktorov operator and the auxiliary state using equation \eqref{eq:aux_state}
\begin{equation} \label{eq:Aux_merging}
f_{m}(\xi)
=
|e^{\xi\,\mathrm{Tr}(\mathbf{z})}|^2|C_\xi|^2
\left|
\langle \boldsymbol{m} |
\hat D(\boldsymbol\beta)\hat R(\mathbf{U_L})\hat S(\boldsymbol r)\hat R(\mathbf{U_R})
\hat D(\bar{\boldsymbol\gamma}_\xi)\hat R(\bar{\mathbf U}_\xi)\hat S(\bar{\boldsymbol\Sigma}_\xi)
|\mathbf 0\rangle
\right|^2.
\end{equation}
It is possible, and convenient to merge all displacement, rotation and squeezing operators together (see Methods section). After merging these operators, the final form of the auxiliary function becomes

\begin{figure}[!ht]
    \centering
    \includegraphics[width = 1\textwidth]{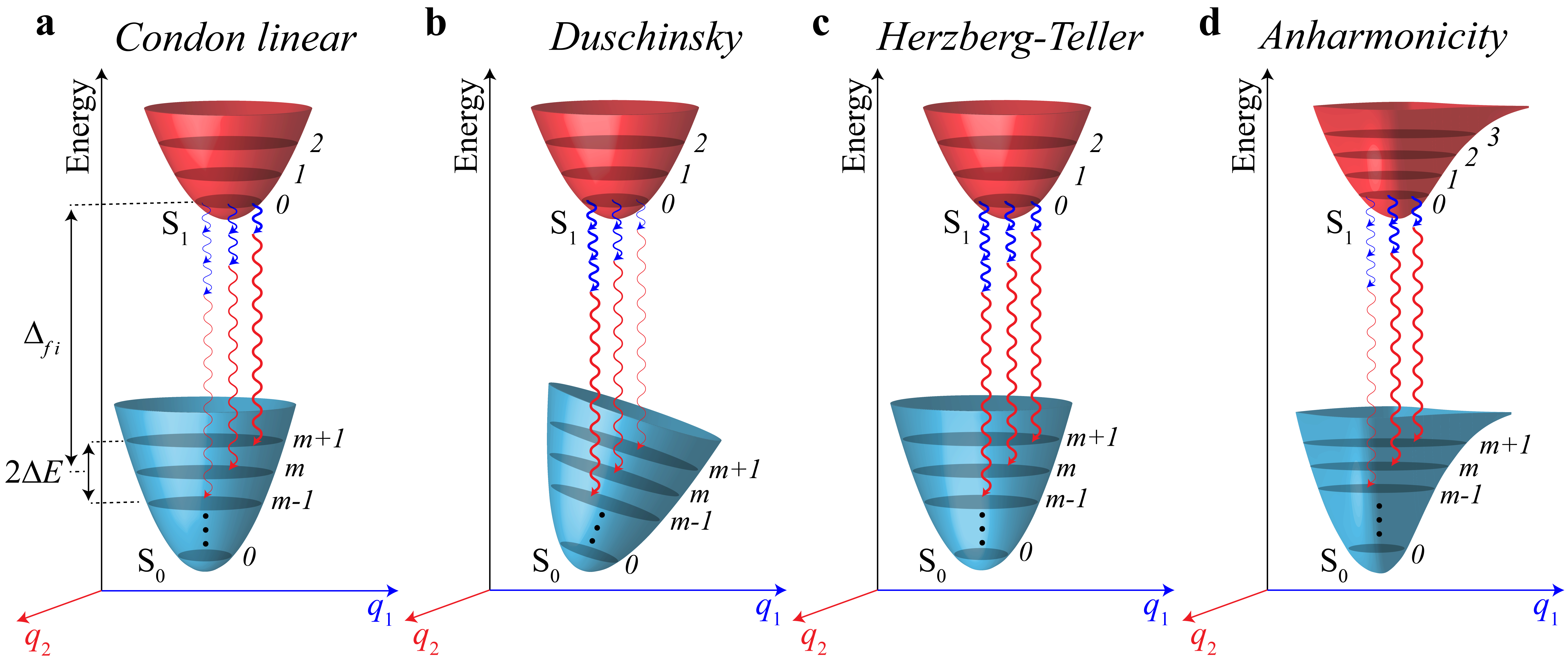}
    \caption{
    Schematic representation of internal conversion process in a two-mode system. (a) Internal conversion occurs from the lowest vibrational level of the $\mathrm{S}_1$ potential energy surface to highly excited vibrational states of $\mathrm{S}_0$ whose total vibrational energy matches the electronic gap $\Delta_{fi}$ within a finite energy window of width $2\Delta E$. Contributions from mode 1 and 2 are highlighted in blue and red colours, respectively. Arrow thickness indicates the relative strength of the transition, determined by the corresponding NACME values. (b) Duschinsky rotation mixes the normal modes and modifies the vibrational wavefunction overlaps. (c) Herzberg-Teller contribution introduces coordinate dependence into the electronic coupling, thereby re-weighting certain non-radiative channels. (d) Anharmonicity modifies the PES shape, vibrational energies, and wavefunctions; unlike the harmonic effects in (a)-(c), it is \textit{not captured} by the standard Gaussian boson sampling operations alone. Further details are discussed in the main text.}
    \label{Fig:IC_concept}
\end{figure}

\begin{equation}\label{eq:Aux_f_final_merge}
f_{m}(\xi)
=
|e^{\xi\,\mathrm{Tr}(\mathbf{z})}|^2|C_\xi|^2
\left|
\left\langle \boldsymbol m\left|
\hat D(\boldsymbol\beta_\xi)\,
\hat R(\mathbf U_\xi)\,
\hat S(\boldsymbol\Sigma_\xi)\,
\right|\mathbf 0\right\rangle
\right|^2.
\end{equation}

The equations \eqref{eq:Superposition} and \eqref{eq:Aux_f_final_merge} constitute the full mapping of $k_\mathrm{IC}$ onto the GBS-compatible sampling procedure that incorporates Duschinsky rotations and HT contribution. The experimental implementation of this framework entails four independent sampling experiments for each term in equation \eqref{eq:Superposition}, as schematically shown in Figure \ref{Fig:GBS_concept}. The statistical outcome of each sampling experiment is used to estimate the square of the corresponding NACME values, which then enter the $k_\mathrm{IC}$ calculation. Moreover, each photon-number-resolved pattern corresponds to the excitation of a specific set of vibrational modes in a molecule, enabling the detailed vibrational mode-resolved analysis.

\subsection*{Physical effects contributing to $k_\mathrm{IC}$}

The model represented by equations \eqref{eq:Superposition} and \eqref{eq:Aux_f_final_merge} combines several physical effects, whose general mechanisms in shaping $k_\mathrm{IC}$ are discussed below. To build intuition, we first consider a simple two-mode system. In the Condon case under the linear coupling approximation, where
$\hat U_{Dok}\hat V=\hat D(\boldsymbol{\beta}_{eff})$, the normal modes behave as independent harmonic oscillators, and the effective displacement $\boldsymbol{\beta}_{eff}$ determines the NACME values as shown in Figure \ref{Fig:IC_concept}(a). Mode 1, indicated by the blue wavy arrows, has relatively low vibrational energy; therefore, satisfying the energy-gap condition $\Delta_{fi}\pm \Delta E$ using this mode alone is extremely unlikely, making this channel weak. By contrast, mode 2, indicated by the red wavy arrows, carries higher vibrational energy, thus combinations involving mode 2 and additional quanta of mode 1 can satisfy the energy-gap condition more efficiently. The channel strength is indicated by the arrow thickness, determined by the corresponding NACME values.

Duschinsky rotations modify this independent-mode picture by rotating the principal axes of the potential energy surfaces (PESs), as shown in Figure \ref{Fig:IC_concept}(b). As a result, the wavefunction overlaps no longer factorize into independent contributions from modes 1 and 2, allowing certain non-radiative channels to be enhanced or suppressed. In contrast, the HT effect does not change the shape or orientation of the PESs, but introduces coordinate dependence into the electronic coupling and thereby re-weights different vibronic channels, as shown in  \ref{Fig:IC_concept}(c). Finally, anharmonicity goes beyond the quadratic PES approximation, since it modifies the PES shape, vibrational energy levels, wavefunctions, and consequently the NACME values (see Figure \ref{Fig:IC_concept}(d)). Anharmonicity introduces non-Gaussian features that cannot be captured by standard GBS operations alone and therefore remains outside the present Gaussian framework.

Different combinations of these physical effects -- except for anharmonicity -- define a hierarchy of approximations within our model. We use these approximations to study their impact on $k_\mathrm{IC}$ values, as discussed below. The corresponding auxiliary functions for these approximations are given in the Methods section.

\subsection*{Sampling results}

\begin{table*}
\centering
\caption{Comparison of IC rate constants obtained using different approximations via sampling (Samp.) and the dynamic programming (DP) method from quantum chemistry \cite{ValievDBT}. The Condon linear coupling approximation is associated with sampling from Poisson distribution, whereas all other cases correspond to GBS approach. The standard deviation $\sigma$ is calculated from the output sampling data using the error propagation method. The standard deviation is not available (NA) for the DP and the experimental values. The Duschinsky mode-mixing effect in the DP method was accounted for using the perturbation theory (see Methods section). Experimentally measured values for $k_\mathrm{IC}$ are specified for the energy gap of 12050 -- 13450 cm$^{-1}$\cite{MishraDBT2025}.}
\label{tab:dbt_comparison}

\small
\setlength{\tabcolsep}{4.5pt}
\renewcommand{\arraystretch}{1.0}

\begin{threeparttable}
\begin{tabular}{lccccccccc}
\toprule
\rowcolor{NatureHeader}
&
\multicolumn{2}{c}{\makecell{Condon\\linear coupling}} &
\multicolumn{2}{c}{\makecell{Condon with\\Duschinsky}} &
\multicolumn{2}{c}{Herzberg-Teller} &
\multicolumn{2}{c}{\makecell{Herzberg-Teller\\with Duschinsky}} &
\multicolumn{1}{c}{\makecell{Measured\\values \cite{MishraDBT2025}}} \\
\cmidrule(lr){2-3}
\cmidrule(lr){4-5}
\cmidrule(lr){6-7}
\cmidrule(lr){8-9}
\cmidrule(lr){10-10}

\rowcolor{NatureHeader}
Quantity
& Samp. & DP
& Samp. & DP
& Samp. & DP
& Samp. & DP 
& Experiment\\
\midrule

$k_\mathrm{IC}$ (s$^{-1}$)
& $2.3\cdot 10^{6}$ & $2.4\cdot 10^{6}$
& $3.7\cdot 10^{7}$ & $6.6\cdot 10^{6}$
& $4.8\cdot 10^{7}$ & $4.9\cdot 10^{7}$
& $2.4\cdot 10^{8}$ & ${5.0\cdot 10^{7}}$
& $1.1 - 7.8\cdot 10^{8}$\\
\rowcolor{NatureStripe}
$\sigma$ (s$^{-1}$)
& $0.4\cdot 10^{6}$ & NA
& $1.0\cdot 10^{7}$ & NA
& $0.2\cdot 10^{7}$ & NA
& $0.1\cdot 10^{8}$ & NA
& NA\\

\bottomrule
\end{tabular}

\end{threeparttable}
\end{table*}

First, we benchmark the $k_\mathrm{IC}$ values obtained via sampling under different levels of approximation against the reference data from the dynamic programming (DP) method known in quantum chemistry \cite{ValievDBT}, the results for which are shown in Table \ref{tab:dbt_comparison}. The detailed description of the sampling simulations as well as the exact procedure for calculating $k_\mathrm{IC}$ values can be found in the Methods section. It is important to note that $k_\mathrm{IC}$ is sensitive to the input parameters from quantum chemistry, meaning that agreement at the order-of-magnitude level provides a meaningful comparison of the methods \cite{ValievDBT}. Within this expected level of accuracy, the GBS and DP results are in excellent agreement for the Condon and HT approximations \textit{without} Duschinsky rotations, as can be seen in Table \ref{tab:dbt_comparison}.

The discrepancy between the methods is observed for the cases when the Duschinsky rotations are included. This difference arises because the perturbation theory used in the DP method assumes that the Duschinsky rotations are negligible, which may not be fully valid for DBT (see Methods section for more details). On the contrary, our GBS-based framework accounts for Duschinsky rotations explicitly, making the results potentially more accurate. Importantly, the $k_\mathrm{IC}$ value obtained from the HT approximation with Duschinsky rotations $2.4\cdot 10^{8}$ s$^{-1}$ matches the experimentally measured value for the DBT molecule  $1.1 - 7.8\cdot 10^{8}$ s$^{-1}$ within the energy gap of 12050 -- 13450 cm$^{-1}$\cite{MishraDBT2025}. However, quantum chemistry methods which include anharmonic effects without Duschinsky rotations \cite{ValievDBT} also reproduce the experimental values, yielding $k_\mathrm{IC}=3.2\cdot 10^{8}$ s$^{-1}$. This indicates that anharmonicity remains an important ingredient for accurate $k_\mathrm{IC}$ calculations.

\begin{table*}[p]
\centering
\caption{Mode-resolved contributions to $k_\mathrm{IC}$ from C--H mode family at 12500 cm$^{-1}$ energy gap for (a) Herzberg-Teller approximation without Duschinsky rotations and (b) Herzberg-Teller approximation with Duschinsky rotations. The table includes C--H mode number, its vibrational frequency $\omega_\mathrm{CH}$, total number of excitations of a C--H mode as well as the number of occurrences of this mode in four terms in equation \eqref{eq:Superposition}. Single and double indicate the number of times a C--H mode was excited ones or twice, respectively. Statistics of both tables are calculated from sampling simulations with 1 billion samples per term in equation \eqref{eq:Superposition}. C--H modes with the highest contributions to $k_\mathrm{IC}$ are highlighted in bold. Note that rows are \textit{not mutually exclusive}, meaning that one pattern $\boldsymbol{m}$ with multiple excited C--H modes contributes to several rows simultaneously.}

\label{xh_mode_contributions}

\scriptsize
\setlength{\tabcolsep}{5.2pt}
\renewcommand{\arraystretch}{1.12}

\vspace{1mm}

\textbf{(a)} Herzberg-Teller approximation without Duschinsky rotations

\vspace{1mm}

\begin{tabular}{cccccccccc}
\toprule
\rowcolor{NatureHeader} \begin{tabular}[c]{@{}c@{}}C--H mode\\ number\end{tabular} & $\omega_{\mathrm{CH}}$ (cm$^{-1}$) & \begin{tabular}[c]{@{}c@{}}Total\\ excitations\end{tabular} & $f_m(0)$ & $f_m(+\tau)$ & $f_m(-\tau)$ & $f_m(i\tau)$ & Single & Double & \multicolumn{1}{c}{$k_{\mathrm{IC}}\pm \sigma$ (s$^{-1}$)} \\
\midrule
$\boldsymbol{149}$ & 3218.122 & 191 & 0 & 67 & 69 & 55 & 189 & 2 & $\boldsymbol{(1.1\pm0.1)\cdot 10^{7}}$ \\
\rowcolor{NatureStripe}
$\boldsymbol{150}$ & 3218.157 & 182 & 0 & 61 & 64 & 57 & 181 & 1 & $\boldsymbol{(1.0\pm0.1)\cdot 10^{7}}$ \\
$\boldsymbol{151}$ & 3218.481 & 162 & 0 & 60 & 53 & 49 & 159 & 3 & $\boldsymbol{(0.9\pm0.1)\cdot 10^{7}}$ \\
\rowcolor{NatureStripe}
$\boldsymbol{152}$ & 3218.501 & 218 & 4 & 76 & 74 & 64 & 201 & 17 & $\boldsymbol{(1.1\pm0.1)\cdot 10^{7}}$ \\
153 & 3225.066 & 42 & 0 & 12 & 17 & 13 & 41 & 1 & $(2.4\pm0.4)\cdot 10^{6}$ \\
\rowcolor{NatureStripe}
154 & 3225.118 & 15 & 0 & 6 & 5 & 4 & 15 & 0 & $(0.8\pm0.2)\cdot 10^{6}$ \\
155 & 3235.112 & 89 & 0 & 33 & 31 & 25 & 86 & 3 & $(0.5\pm0.1)\cdot 10^{7}$ \\
\rowcolor{NatureStripe}
156 & 3235.234 & 89 & 0 & 34 & 32 & 23 & 87 & 2 & $(0.5\pm0.1)\cdot 10^{7}$ \\
$\boldsymbol{157}$ & 3235.286 & 120 & 0 & 41 & 45 & 34 & 117 & 3 & $\boldsymbol{(0.7\pm0.1)\cdot 10^{7}}$ \\
\rowcolor{NatureStripe}
158 & 3235.288 & 77 & 0 & 26 & 29 & 22 & 75 & 2 & $(0.4\pm0.1)\cdot 10^{7}$ \\
159 & 3239.330 & 18 & 0 & 6 & 4 & 8 & 18 & 0 & $(1.1\pm0.3)\cdot 10^{6}$ \\
\rowcolor{NatureStripe}
160 & 3239.437 & 45 & 2 & 18 & 10 & 15 & 43 & 2 & $(2.2\pm0.5)\cdot 10^{6}$ \\
161 & 3248.899 & 40 & 0 & 12 & 16 & 12 & 38 & 2 & $(2.3\pm0.4)\cdot 10^{6}$ \\
\rowcolor{NatureStripe}
$\boldsymbol{162}$ & 3249.121 & 103 & 0 & 37 & 35 & 31 & 98 & 5 & $\boldsymbol{(0.6\pm0.1)\cdot 10^{7}}$ \\
163 & 3249.222 & 29 & 0 & 7 & 12 & 10 & 28 & 1 & $(1.7\pm0.3)\cdot 10^{6}$ \\
\rowcolor{NatureStripe}
164 & 3249.469 & 68 & 0 & 30 & 18 & 20 & 65 & 3 & $(3.8\pm0.5)\cdot 10^{6}$ \\
165 & 3262.950 & 82 & 0 & 33 & 20 & 29 & 80 & 2 & $(0.5\pm0.1)\cdot 10^{7}$ \\
\rowcolor{NatureStripe}
166 & 3262.970 & 57 & 0 & 17 & 17 & 23 & 57 & 0 & $(3.5\pm0.5)\cdot 10^{6}$ \\
167 & 3265.036 & 45 & 0 & 16 & 16 & 13 & 45 & 0 & $(2.5\pm0.4)\cdot 10^{6}$ \\
\rowcolor{NatureStripe}
$\boldsymbol{168}$ & 3265.131 & 114 & 2 & 34 & 43 & 35 & 111 & 3 & $\boldsymbol{(0.6\pm0.1)\cdot 10^{7}}$ \\
\bottomrule
\end{tabular}

\scriptsize
\setlength{\tabcolsep}{5.2pt}
\renewcommand{\arraystretch}{1.12}

\vspace{4mm}

\textbf{(b)} Herzberg-Teller approximation with Duschinsky rotations

\vspace{1mm}

\begin{tabular}{cccccccccc}
\toprule
\rowcolor{NatureHeader} \begin{tabular}[c]{@{}c@{}}C--H mode\\ number\end{tabular} & $\omega_{\mathrm{CH}}$ (cm$^{-1}$) & \begin{tabular}[c]{@{}c@{}}Total\\ excitations\end{tabular} & $f_m(0)$ & $f_m(+\tau)$ & $f_m(-\tau)$ & $f_m(i\tau)$ & Single & Double & \multicolumn{1}{c}{$k_{\mathrm{IC}}\pm \sigma$ (s$^{-1}$)} \\
\midrule
$\boldsymbol{149}$ & 3218.122 & 1817 & 273 & 520 & 509 & 515 & 1696 & 118 & $\boldsymbol{(4.3\pm0.4)\cdot 10^{7}}$ \\
\rowcolor{NatureStripe}
$\boldsymbol{150}$ & 3218.157 & 1460 & 181 & 409 & 439 & 431 & 1396 & 63 & $\boldsymbol{(4.3\pm0.3)\cdot 10^{7}}$ \\
$\boldsymbol{151}$ & 3218.481 & 1133 & 138 & 301 & 337 & 357 & 1100 & 31 & $\boldsymbol{(3.5\pm0.3)\cdot 10^{7}}$ \\
\rowcolor{NatureStripe}
$\boldsymbol{152}$ & 3218.501 & 1281 & 158 & 347 & 394 & 382 & 1228 & 53 & $\boldsymbol{(3.8\pm0.3)\cdot 10^{7}}$ \\
153 & 3225.066 & 577 & 68 & 148 & 203 & 158 & 563 & 14 & $(1.7\pm0.2)\cdot 10^{7}$ \\
\rowcolor{NatureStripe}
154 & 3225.118 & 348 & 53 & 113 & 89 & 93 & 340 & 8 & $(0.8\pm0.2)\cdot 10^{7}$ \\
$\boldsymbol{155}$ & 3235.112 & 2346 & 446 & 609 & 655 & 636 & 2067 & 267 & $\boldsymbol{(3.4\pm0.5)\cdot 10^{7}}$ \\
\rowcolor{NatureStripe}
156 & 3235.234 & 465 & 43 & 127 & 147 & 148 & 463 & 2 & $(1.7\pm0.2)\cdot 10^{7}$ \\
$\boldsymbol{157}$ & 3235.286 & 1115 & 115 & 287 & 356 & 357 & 1063 & 51 & $\boldsymbol{(3.9\pm0.3)\cdot 10^{7}}$ \\
\rowcolor{NatureStripe}
$\boldsymbol{158}$ & 3235.288 & 2253 & 430 & 608 & 608 & 607 & 1978 & 256 & $\boldsymbol{(3.2\pm0.4)\cdot 10^{7}}$ \\
159 & 3239.330 & 1639 & 334 & 410 & 488 & 407 & 1516 & 122 & $(1.7\pm0.4)\cdot 10^{7}$ \\
\rowcolor{NatureStripe}
160 & 3239.437 & 1357 & 290 & 328 & 384 & 355 & 1267 & 86 & $(1.2\pm0.4)\cdot 10^{7}$ \\
161 & 3248.899 & 545 & 65 & 147 & 176 & 157 & 530 & 14 & $(1.7\pm0.2)\cdot 10^{7}$ \\
\rowcolor{NatureStripe}
162 & 3249.121 & 977 & 113 & 285 & 289 & 290 & 946 & 30 & $(3.1\pm0.3)\cdot 10^{7}$ \\
163 & 3249.222 & 1015 & 186 & 289 & 264 & 276 & 974 & 41 & $(1.6\pm0.3)\cdot 10^{7}$ \\
\rowcolor{NatureStripe}
164 & 3249.469 & 1167 & 224 & 307 & 336 & 300 & 1114 & 51 & $(1.6\pm0.3)\cdot 10^{7}$ \\
165 & 3262.950 & 1241 & 224 & 339 & 334 & 344 & 1157 & 83 & $(2.1\pm0.3)\cdot 10^{7}$ \\
\rowcolor{NatureStripe}
166 & 3262.970 & 1291 & 242 & 311 & 370 & 368 & 1201 & 89 & $(2.0\pm0.3)\cdot 10^{7}$ \\
167 & 3265.036 & 676 & 122 & 174 & 191 & 189 & 651 & 25 & $(1.1\pm0.2)\cdot 10^{7}$ \\
\rowcolor{NatureStripe}
168 & 3265.131 & 1230 & 220 & 339 & 327 & 344 & 1192 & 37 & $(2.1\pm0.3)\cdot 10^{7}$ \\
\bottomrule
\end{tabular}

\end{table*}


The advantage of the present GBS framework is that the sampling data retain explicit information about excitations in each normal mode, enabling the detailed mode-resolved analysis of $k_\mathrm{IC}$. As a first step, however, it is important to evaluate the involvement of different mode families in the IC process of DBT. This molecule has 168 normal modes, which can be divided into the mode family categories -- namely low-frequency C--C modes $\omega=\{0-1000\}$ cm$^{-1}$, high-frequency C--C modes $\omega=\{1000-3000\}$ cm$^{-1}$, and C--H modes $\omega>3000$ cm$^{-1}$ -- with the corresponding mode numbers 1--81, 82--148 and 149--168, respectively. We find that the C--H mode family dominates $k_\mathrm{IC}$, being involved in more than $>95\%$ of the energy-window photon patterns and accounting for more than $>99\%$ of the total rate contribution. Therefore we perform the detailed mode-resolved analysis specifically for this mode family.

\begin{figure}[!ht]
    \centering
    \includegraphics[width = 1\textwidth]{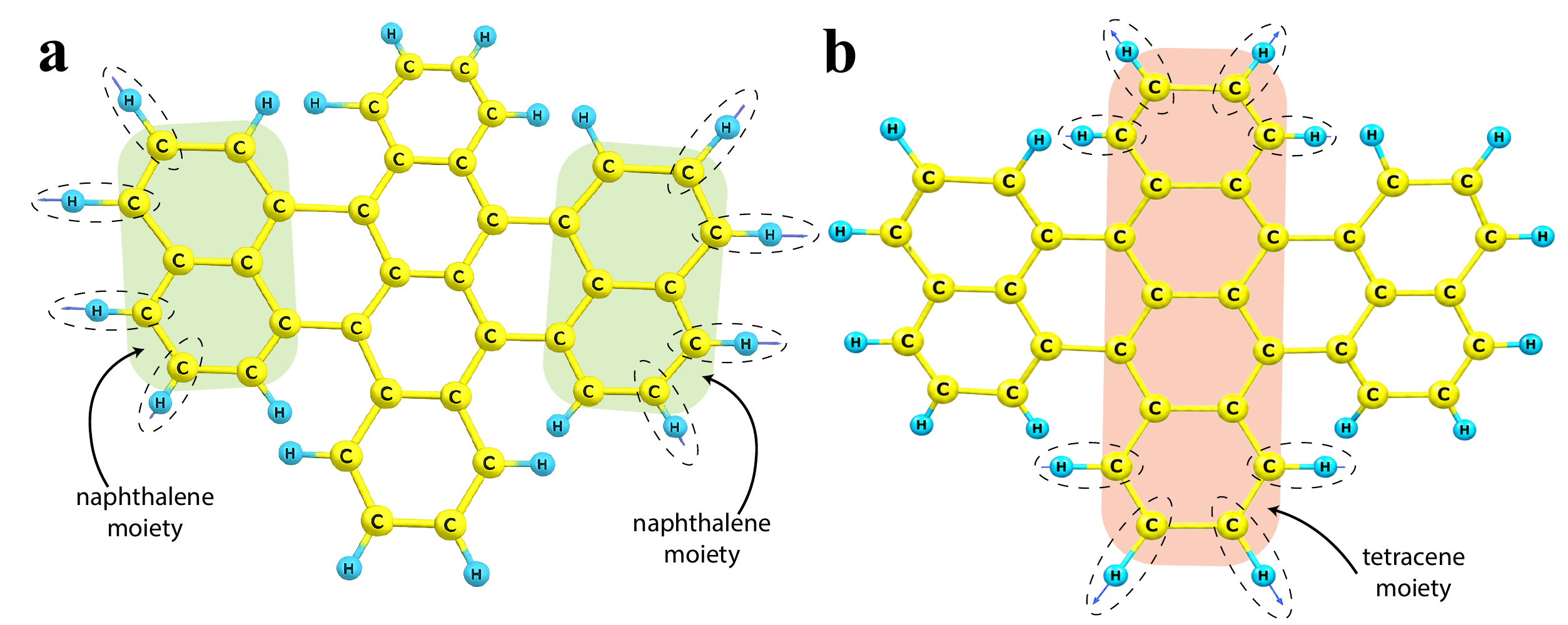}
    \caption{
     Graphical representation of the DBT molecule. (a) The C--H bonds involved in the vibrations of normal modes 149-152 are encircled. The corresponding modes are located on the naphthalene moiety of DBT, highlighted in the olive colour. (b) The C--H bonds corresponding to normal modes 158 and 159 are encircled. These modes are located on the tetracene moiety of DBT, highlighted in the pink colour. We note that the distinct C--H \textit{bonds} contribute to vibrations of \textit{multiple normal modes}, which are depicted as blue arrows near the Hydrogen atoms.}
    \label{Fig:Results}
\end{figure}

It is especially informative to compare the results of the C--H mode-resolved analysis between the HT approximation with and without Duschinsky rotations. For the no-Duschinsky case (see Table \ref{xh_mode_contributions}), the total number of C--H excitations remains comparatively small, especially for the $f_m(0)$ term, where almost no C--H excitations are observed. The physical origin of this behavior lies in the extremely small displacement parameters $\boldsymbol{\beta}$ of the C--H modes. As a consequence, they are highly unlikely to be excited without the impact arising from the NAC operator. On the contrary, the terms affected by the NAC operator -- namely $f_m(+\tau)$, $f_m(-\tau)$ and $f_m(i\tau)$ -- show a high probability of excitation. The individual $k_\mathrm{IC}$ contributions shown in Table \ref{xh_mode_contributions} are obtained by summing the NACME values for all photon patterns in which a particular mode is excited. With this metric, the most dominant C--H modes for the no-Duschinsky case are 149--152, which are all located on the naphthalene moieties of DBT \cite{ValievDBT} highlighted in Figure \ref{Fig:Results}.

When Duschinsky rotations are included, the total number of unique patterns with C--H excitations increases substantially -- from 858 to 13638 -- leading to larger contributions from all C--H modes (see Table \ref{xh_mode_contributions}(b)). Notably, modes with the highest contributions to $k_\mathrm{IC}$ -- 149--152 -- remain the same in both scenarios. However, certain C--H modes such as 155 and 159 show close to an order-of-magnitude enhancement in the $k_\mathrm{IC}$ contributions when the Duschinsky rotations are included. This raises an important question: what causes such selective enhancement of certain modes in the Duschinsky model?

\begin{figure}[!t]
    \centering
    \includegraphics[width = 0.6\textwidth]{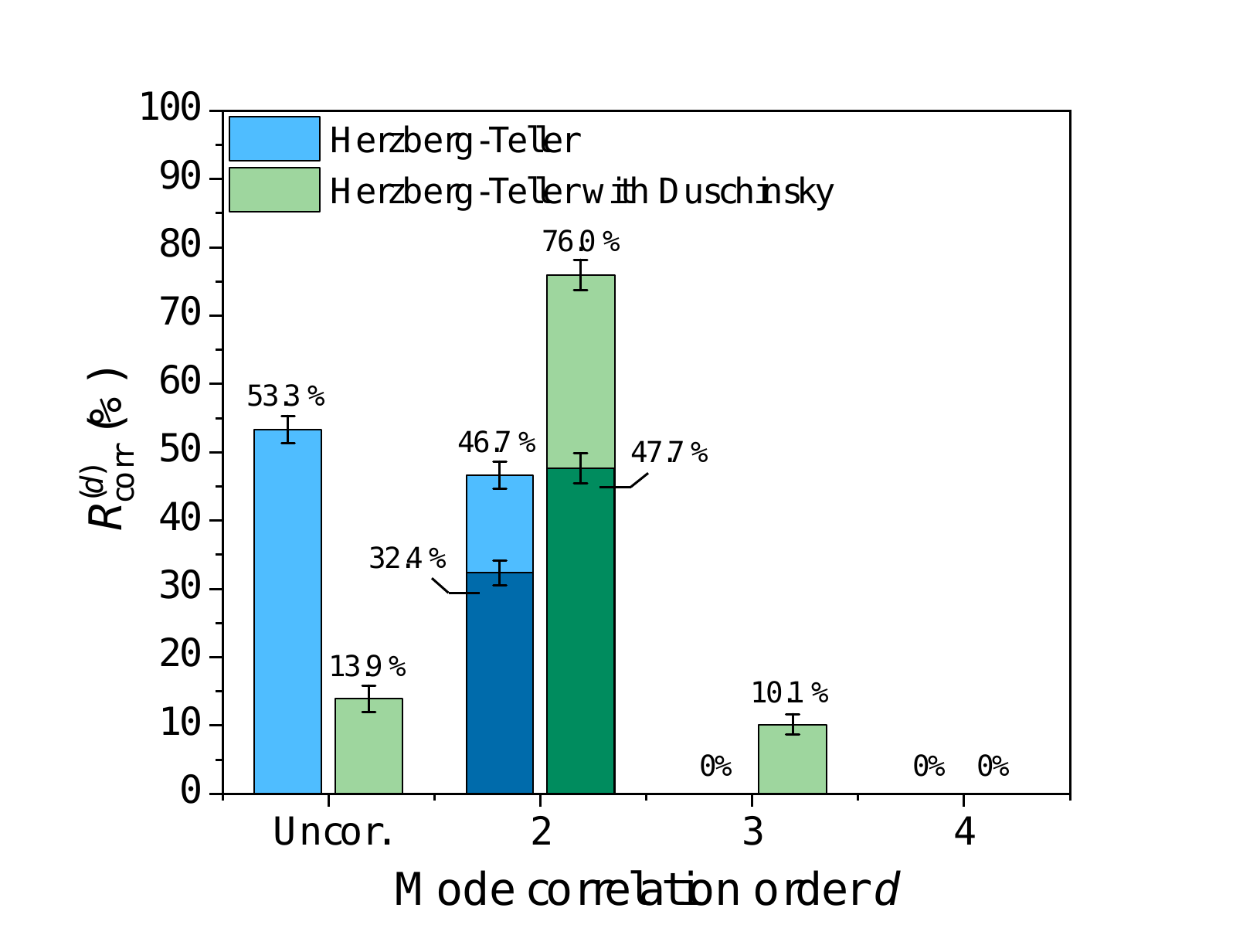}
    \caption{Exclusive correlated rate fraction $R_\mathrm{corr}^{(d)}(F)$ across all pathway fragments $F$ with respect to the mode correlation order $d$ for the HT model with and without Duschinsky rotations. Darkened regions on the bars of second order correspond to the highly-correlated rate fractions with $g_{\mathrm{occ}}^{(2)}(F)>2$.}
    \label{Fig:correlations}
\end{figure}

To answer this question, we consider the co-occurrence events -- a scenario in which several modes are excited simultaneously within the same vibrational pathway. A recurring set of $d$ distinct modes is referred to as a pathway fragment $F=\{i_1,\ldots,i_d\}$, where $d=|F|$ defines its correlation order. For each photon pattern $\boldsymbol{m}$, the occupation of mode $i$ is represented by the binary variable $b_i$, with $b_i=1$ for $m_i>0$ and $b_i=0$ for $m_i=0$. Considering only patterns in the energy window $\boldsymbol{m}\in Q_{\Delta E}$, we define
\begin{equation}
P(F)
=
P\!\left(
b_i=1\ \text{for all}\ i\in F
\,\middle|\,
\boldsymbol{m}\in Q_{\Delta E}
\right),
\end{equation}
as the probability that all vibrational modes in $F$ occur within a pathway $\boldsymbol{m}$. To distinguish higher-order correlations from co-occurrences inherited from lower-order fragments, we define the normalized co-occurrence factor as
\begin{equation}
g_{\mathrm{occ}}^{(d)}(F)
=
\min_{S\mid(F\setminus S)}
\frac{
P(F)
}{
P(S)\,
P(F\setminus S)
},
\label{eq:binary_gd}
\end{equation}
where $S$ denotes a non-empty subfragment of $F$ with the minimum being evaluated over all unique bipartitions of $F$. Values of $g_{\mathrm{occ}}^{(d)}(F)>1$ indicate that the complete pathway fragment occurs more frequently than predicted by each of its possible lower-order bipartitions. Therefore, we attribute a fragment $F$ to the correlated pathways when its lower bound exceeds the threshold value $g_{\mathrm{occ}}^{(d)}(F)>1.0$, while fragments that do not pass this threshold are treated as uncorrelated. Fragments with $g_{\mathrm{occ}}^{(d)}(F)>2$ are considered highly correlated, since their joint occurrence probability is at least twice higher than the expectation for every bipartition. The total rate fraction associated with a fragment $F$ is
\begin{equation}
R_\mathrm{corr}(F)
=
\frac{k_{\mathrm{IC}}(F)}{k_{\mathrm{IC}}},
\qquad
k_{\mathrm{IC}}(F)
=
\sum_{\boldsymbol m\supset F} k_{\mathrm{IC}}(\boldsymbol m),
\end{equation}
where $\boldsymbol m\supset F$ denotes a pathway containing all modes of $F$. Since a pathway may contain fragments of several orders simultaneously, adding their individual rate fractions would result in double counting, therefore we assign each pathway to the highest available fragment order. Denoting the total rate assigned to order $d$ by $k_{\mathrm{IC}}^{(d)}(F)$, the corresponding exclusive rate fraction is
\begin{equation}
R_\mathrm{corr}^{(d)}(F)
=
\frac{k_{\mathrm{IC}}^{(d)}(F)}{k_{\mathrm{IC}}}.
\end{equation}

We observe that in the no-Duschinsky model 46.7\% of the total rate is associated with resolved correlated pathways, whereas Duschinsky rotations increase this fraction to 86.1\% (see Figure \ref{Fig:correlations}). In the Duschinsky model, second- and third-order correlations account for 76.0\% and 10.1\% of $k_\mathrm{IC}$, respectively. Highly correlated fragments, defined by $g_{\mathrm{occ}}^{(d)}(F)>2$, occur exclusively at second order and collectively contribute 47.7\% and 32.4\% of the rate with and without Duschinsky rotations, respectively. Table \ref{xh_mode_contributions} shows these highly-correlated fragments with the largest individual rate contributions. These results demonstrate that without Duschinsky rotations, the correlated fraction is carried by only three second-order fragments, whereas with Duschinsky rotations it is distributed over many different fragments. This highlights the importance of Duschinsky rotations, which organize a large part of $k_\mathrm{IC}$ into distinct low-order fragments, while without it the rate remains mostly pathway-delocalized.

\begin{table}[t]
\centering
\caption{
Highly correlated pathway fragments $F$ with the largest $k_\mathrm{IC}$ contribution obtained for the HT model with and without Duschinsky rotations. Pathway fragments are considered highly correlated when $g_{\mathrm{occ}}^{(d)}(F)>2$, which is observed only for the second correlation order. The rate contribution $R_\mathrm{corr}(F)$ is non-exclusive, meaning that pathways containing multiple fragments may contribute to more than one entry.}
\label{tab:correlated_fragments}

\renewcommand{\arraystretch}{1.4}
\begin{tabular}{lcc@{\hspace{2em}}lcc}
\toprule
\rowcolor{NatureHeader}
\multicolumn{3}{c}{\textbf{HT without Duschinsky rotations}}
&
\multicolumn{3}{c}{\textbf{HT with Duschinsky rotations}} \\
\midrule

\rowcolor{NatureHeader}
Fragment $F$
&
$g_{\mathrm{occ}}^{(2)}(F)$
&
$R_\mathrm{corr}^{(2)}(F)$
&
Fragment $F$
&
$g_{\mathrm{occ}}^{(2)}(F)$
&
$R_\mathrm{corr}^{(2)}(F)$ \\
\midrule
$\{149,150\}$ & $2.98_{-0.17}^{+0.18}$ & $14.0 \pm 1.3\,\%$ & $\{149,150\}$ & $2.69_{-0.09}^{+0.09}$ & $10.1 \pm 0.7\,\%$ \\
\rowcolor{NatureStripe}
$\{151,152\}$ & $2.16_{-0.14}^{+0.16}$ & $10.0 \pm 1.1\,\%$ & $\{152,157\}$ & $3.09_{-0.13}^{+0.14}$ & $6.8 \pm 0.5\,\%$ \\
$\{152,157\}$ & $2.63_{-0.18}^{+0.19}$ & $8.8 \pm 1.0\,\%$ & $\{151,152\}$ & $2.51_{-0.13}^{+0.13}$ & $5.6 \pm 0.5\,\%$ \\
\rowcolor{NatureStripe}
NA & NA & NA & $\{3,155\}$ & $3.30_{-0.09}^{+0.10}$ & $4.5 \pm 1.0\,\%$ \\
NA & NA & NA & $\{151,156\}$ & $4.24_{-0.26}^{+0.27}$ & $3.7 \pm 0.4\,\%$ \\
\rowcolor{NatureStripe}
NA & NA & NA & $\{74,162\}$ & $3.54_{-0.22}^{+0.24}$ & $3.3 \pm 0.4\,\%$ \\
NA & NA & NA & $\{166,167\}$ & $2.75_{-0.19}^{+0.20}$ & $2.8 \pm 0.3\,\%$ \\
\rowcolor{NatureStripe}
NA & NA & NA & $\{101,168\}$ & $2.31_{-0.09}^{+0.10}$ & $2.8 \pm 0.8\,\%$ \\
NA & NA & NA & $\{61,152\}$ & $2.02_{-0.12}^{+0.13}$ & $2.2 \pm 0.5\,\%$ \\
\rowcolor{NatureStripe}
NA & NA & NA & $\{4,130\}$ & $8.21_{-0.39}^{+0.41}$ & $2.0 \pm 0.6\,\%$ \\
\bottomrule
\end{tabular}

\end{table}

\section*{Conclusion and Discussion}

In this work, we map $k_\mathrm{IC}$ calculations onto a GBS-compatible sampling procedure incorporating the HT contribution and Duschinsky rotations and show that our approach reproduces $k_\mathrm{IC}$ values of DBT obtained in previous studies \cite{MishraDBT2025,ValievDBT}. Moreover, unlike standard methods in chemistry, which treat $k_\mathrm{IC}$ as a single aggregated value, our framework decomposes the rate constant $k_\mathrm{IC}$ into contributions from distinct vibrational pathways associated with the corresponding photon-number patterns in GBS. This decomposition reveals correlations between vibrational modes, showing that in the presence of Duschinsky rotations, more than 86\% of the rate is organized into recurring low-order mode combinations. Such insights can guide targeted molecular modifications, including selective deuteration \cite{Valiev2020} or bond substitution, to suppress or enhance specific non-radiative pathways.

A natural extension of the presented method is the inclusion of finite-temperature effects, which have previously been incorporated into simulations of vibronic spectra \cite{Huh2017VibronicGBS}. Such an extended GBS mapping could be applied to other processes governed by NAC type of coupling operators, including multiphonon non-radiative recombination in solids \cite{Turiansky2021}. A further important direction is the treatment of anharmonicity, although its inclusion would require extending the framework beyond the capabilities of standard GBS \cite{Wang2022Hybrid}, which may provide a path forward for demonstrating quantum advantage.

To illustrate the associated computational demands, we note that generating the required four billion samples in our \textit{classical} sampling simulations takes approximately 10 hours of parallel computation on a high-performance computing cluster. Recent experimental demonstration of GBS \cite{Stefszky2025PaQS}, however, operates at an effective sampling rate of 1 MHz, which would generate the same number of samples in around one hour, provided that such a system could be scaled to the required size \cite{Liu2026Jiuzhang4}. While this estimate neglects experimental imperfections, it highlights the practical appeal of developing photonic hardware to generate the photon-pattern-resolved information that may otherwise be computationally expensive to produce classically.


\section*{Data Availability}
The molecular and optical parameters, together with the simulated and measured data supporting the findings of this study, will be made publicly available in Zenodo upon publication.

\section*{Code Availability}
The code used to simulate the internal conversion rate constant using Gaussian boson sampling will be made publicly available in Zenodo upon publication.

\section*{Acknowledgments}
The authors acknowledge the computing time made available to them on the high-performance computer (Noctua2 and Otus) at the NHR Center Paderborn Center for Parallel Computing (PC2). This center is jointly supported by the Federal Ministry of Research, Technology and Space and the state governments participating in the National High-Performance Computing (NHR) joint funding program (www.nhr-verein.de/en/our-partners). M.R. would like to thank Robert Schade and Manoar Hossain from PC2 for assistance in setting up simulations on the computing cluster. Authors thank Fabian Schlue and Denis Kopylov for valuable comments on the manuscript. R.V. acknowledges the Academy of Finland (project: 346369). This work has received funding from the German Federal Ministry of Research, Technology and Space (BMFTR) within the PhoQuant project (Grant No. 13N16103).

\section*{Author Contributions}
M.R. and R.V. conceived the presented idea. M.R. derived formulas for mapping IC rate constant calculation onto GBS architecture. M.R. wrote the Python scripts and performed GBS simulations on the computing cluster. R.V. chose the molecular object and provided the input quantum chemistry parameters for it. R.V. provided the reference IC rate constant data used for benchmarking. M.R. and R.V. analyzed the data with input from J.E.. M.R. wrote the manuscript with input from other authors. C.S., M.S., B.B. and T.K. supervised the research. All authors edited the paper.

\section*{Competing Interests}
The authors declare no competing financial or non-financial interests.

\bibliographystyle{naturemag.bst}
\bibliography{References}

\section*{Methods}

\subsection*{Multimode operators and nullifier equations}

The multimode displacement, rotation and squeezing  operators are defined as \cite{MaRhodes1990, Jnane2021}
\begin{equation} \label{eq:Disp_op_define}
\hat D(\boldsymbol{\gamma})
=
\exp\!\left(
\boldsymbol{\gamma}^T \hat{\mathbf a}^\dagger
-
\boldsymbol{\gamma}^\dagger \hat{\mathbf a}
\right),
\end{equation}
\begin{equation} \label{eq:Rotation_op_define}
\hat R(\mathbf U)
=
\exp\!\left[
\hat{\mathbf a}^{\dagger T}\bigl(\ln \mathbf U\bigr)\hat{\mathbf a}
\right],
\end{equation}
\begin{equation} \label{eq:MMSqueezing_op_define}
\hat S(\boldsymbol{\Xi})
=
\exp\!\left[
\frac12
\left(
\hat{\mathbf a}^{\dagger T}\boldsymbol{\Xi}\hat{\mathbf a}^\dagger
-
\hat{\mathbf a}^T\boldsymbol{\Xi}^*\hat{\mathbf a}
\right)
\right],
\end{equation}
where $\mathbf{a}=(a_1, \dots, a_M)^T$ and $\mathbf{a^\dagger}=(a_1^\dagger, \dots, a_M^\dagger)^T$ are the multi-mode ladder operators \cite{MaRhodes1990}. The action of the displacement, rotation and squeezing operators on the ladder operators is given by \cite{MaRhodes1990}
\begin{equation} \label{eq:Disp_action}
\hat D^\dagger(\boldsymbol{\gamma})\,\hat{\mathbf a}\,\hat D(\boldsymbol{\gamma})
=
\hat{\mathbf a}+\boldsymbol{\gamma},
\end{equation}
\begin{equation} \label{eq:Rotation_action}
\hat R^\dagger(\mathbf U)\,\hat{\mathbf a}\,\hat R(\mathbf U)
=
\mathbf U\,\hat{\mathbf a},
\end{equation}
\begin{equation} \label{eq:Squeezing_action}
\hat S^\dagger(\boldsymbol{\Xi})\,\hat{\mathbf a}\,\hat S(\boldsymbol{\Xi})
=
\mathbf{A}\hat{\mathbf a}
+
\mathbf{B}\hat{\mathbf a}^\dagger.
\end{equation}
Here $\boldsymbol{\gamma}$ is a complex vector, $\mathbf U$ is a unitary, $\boldsymbol{\Xi}$ is a complex symmetric matrix, and matrices $\mathbf{A}$ and $\mathbf{B}$ are the Bogoliubov matrices. A common way to write the Bogoliubov matrices in this case is using the polar decomposition \cite{MaRhodes1990}
\begin{equation} \label{eq:Squeezing_action_MaRhodes}
\hat S^\dagger(\boldsymbol{\Xi})\,\hat{\mathbf a}\,\hat S(\boldsymbol{\Xi})
=
\cosh{(\boldsymbol{r}_{\mathrm{pol}})}\hat{\mathbf a}
+
\sinh{(\boldsymbol{r}_{\mathrm{pol}})}e^{i\boldsymbol \theta}\hat{\mathbf a}^\dagger,
\end{equation}
where $\boldsymbol{\Xi} = \boldsymbol{r}_{\mathrm{pol}}e^{i\boldsymbol{\theta}}$ with $\boldsymbol{r}_{\mathrm{pol}}$ being the positive semi-definite matrix representing the squeezing magnitudes and $e^{i\boldsymbol \theta}$ encodes squeezing phases \cite{MaRhodes1990}.

However, in this work we use an alternative representation of equation \eqref{eq:Squeezing_action_MaRhodes}. Since $\boldsymbol{\Xi}$ is complex symmetric matrix, it must admit the Takagi decomposition $\boldsymbol{\Xi}=\mathbf{W}\boldsymbol{r}\mathbf{W}^T$ \cite{CariolaroTAKAGI}, where $\boldsymbol{r} = diag(r_1, \dots ,r_M)$ is the diagonal matrix of single-mode squeezing strengths (real non-negative values) and $\mathbf{W}$ is a unitary, which represents the basis change that leads to the multimode correlation of the squeezed states. Inserting identity matrix $\mathbf{W}^{\dagger}\mathbf{W}$ into the Takagi decomposition $\boldsymbol{\Xi}=\mathbf{W}\boldsymbol{r}\mathbf{W}^{\dagger}\mathbf{W}\mathbf{W}^T$, the correspondence between the polar and Takagi decompositions becomes clear: $\boldsymbol{r}_{\mathrm{pol}}=\mathbf{W}\boldsymbol{r}\mathbf{W}^{\dagger}$, $e^{i\theta} = \mathbf{W}\mathbf{W}^T$. This leads to the Takagi decomposed form of the Bogoliubov transformation
\begin{equation} \label{eq:Squeezing_action_Takagi}
\hat S^\dagger(\boldsymbol{\Xi})\,\hat{\mathbf a}\,\hat S(\boldsymbol{\Xi})
=
\mathbf{W}\cosh{(\boldsymbol{r})}\mathbf{W}^{\dagger}\hat{\mathbf a}
+
\mathbf{W}\sinh{(\boldsymbol{r})}\mathbf{W}^T\hat{\mathbf a}^\dagger.
\end{equation}

Now we introduce the nullifier equation \cite{LasserNULLIFIERS2018,FunaiNULLIFIERS2025} for the general multimode squeezed state. Starting from the vacuum condition \(\hat{\mathbf a}|\mathbf 0\rangle=0\), one has
\begin{equation} \label{eq:nullifier_squeezing}
0
=
\hat S(\boldsymbol{\Xi})\hat{\mathbf a}|\mathbf 0\rangle
=
\left(\hat S(\boldsymbol{\Xi})\hat{\mathbf a}\hat S^\dagger(\boldsymbol{\Xi})\right)\hat S(\boldsymbol{\Xi})|\mathbf 0\rangle
=
\left(\hat S(\boldsymbol{\Xi})\hat{\mathbf a}\hat S^\dagger(\boldsymbol{\Xi})\right)|\psi\rangle,
\end{equation}
where $|\psi\rangle$ is the squeezed vacuum state. Using $S^\dagger(\boldsymbol{\Xi})=S(\boldsymbol{-\Xi})$ and the fact that Takagi decomposition of the negative matrix is $-\boldsymbol{\Xi}=(i\mathbf{W})\boldsymbol{r}(i\mathbf{W}^T)$, it is straightforward to obtain the action entering equation \eqref{eq:nullifier_squeezing}:
\begin{equation} \label{eq:Squeezing_action_Takagi_reversed}
\hat S(\boldsymbol{\Xi})\hat{\mathbf a}\,\hat S^\dagger(\boldsymbol{\Xi})
=
\mathbf{W}\cosh{(\boldsymbol{r})}\mathbf{W}^{\dagger}\hat{\mathbf a}
-
\mathbf{W}\sinh{(\boldsymbol{r})}\mathbf{W}^T\hat{\mathbf a}^\dagger.
\end{equation}
Hence, we obtain the corresponding nullifier equation
\begin{equation} \label{eq:nullifier_squeezing2}
(\mathbf{W}\cosh{(\boldsymbol{r})}\mathbf{W}^{\dagger}\hat{\mathbf a}
-
\mathbf{W}\sinh{(\boldsymbol{r})}\mathbf{W}^T\hat{\mathbf a}^\dagger)|\psi\rangle = 0.
\end{equation}
Since $\mathbf{W}\cosh{(\boldsymbol{r})}\mathbf{W}^{\dagger}$ is invertible, which follows from $\cosh{(r_j)\ge 1}$, we can write the nullifier equation in the following form:
\begin{equation} \label{eq:nullifier_squeezing4}
\left(
\hat{\mathbf a}
-
\mathbf Z\,\hat{\mathbf a}^\dagger
\right)|\psi\rangle=0,
\qquad
\mathbf Z=\mathbf W\tanh(\boldsymbol r)\mathbf W^T.
\end{equation}

For the operator $\hat S(\boldsymbol{\Xi})$ the Bloch-Messiah decomposition can be written as
\begin{equation}\label{eq:multimode_squeezing_decomposition}
\hat S(\boldsymbol{\Xi}) = \hat R(\mathbf{W})\hat S(\boldsymbol{r}) \hat R^\dagger(\mathbf{W}),
\end{equation}
which using $\hat R^\dagger(\mathbf{W})|\mathbf 0\rangle = |\mathbf 0\rangle$ leads to
\begin{equation}\label{eq:multimode_squeezing_decomposition_2}
|\psi\rangle = \hat S(\boldsymbol{\Xi})|\mathbf 0\rangle = \hat R(\mathbf{W})\hat S(\boldsymbol{r})|\mathbf 0\rangle.
\end{equation}

For the general displaced squeezed multimode state we can write
\begin{equation}\label{eq:multimode_squeezing_decomposition_displaced}
|\psi_{\gamma}\rangle=\hat D(\boldsymbol{\gamma})|\psi\rangle = \hat D(\boldsymbol{\gamma})\hat R(\mathbf{W})\hat S(\boldsymbol{r})|\mathbf 0\rangle.
\end{equation}
To obtain the corresponding nullifier equation in this case, we start from equation \eqref{eq:nullifier_squeezing4} and transform it using the displacement operator in equation \eqref{eq:Disp_op_define}
\begin{equation} \label{eq:nullifier_squeezing_displaced_1}
D(\boldsymbol{\gamma})\left(
\hat{\mathbf a}
-
\mathbf Z\,\hat{\mathbf a}^\dagger
\right)D^{\dagger}(\boldsymbol{\gamma})|\psi_{\gamma}\rangle=0.
\end{equation}
From equation \eqref{eq:Disp_action} it follows that $\hat D(\boldsymbol{\gamma})\,\hat{\mathbf a}\,\hat D^\dagger(\boldsymbol{\gamma}) = \hat{\mathbf{a}}-\gamma$ and $\hat D(\boldsymbol{\gamma})\,\hat{\mathbf a}^\dagger\,\hat D^\dagger(\boldsymbol{\gamma}) = \hat{\mathbf{a}}^\dagger-\gamma^*$, which leads to
\begin{equation} \label{eq:nullifier_squeezing_displaced}
\left(
\hat{\mathbf a}
-
\mathbf Z\,\hat{\mathbf a}^\dagger
+
\boldsymbol \eta
\right)|\psi_{\gamma}\rangle=0,
\end{equation}
where we introduced $\boldsymbol{\eta}=\mathbf Z\gamma^*-\gamma$.

Additionally, we note that in the special case of independent single-mode squeezers when $\boldsymbol{\Xi}=\boldsymbol{r}$, operator $\hat S(\boldsymbol{\Xi})$ becomes a product of independent single-mode squeezers, whose action on the ladder operator reduces to \cite{MaRhodes1990}
\begin{equation} \label{eq:Squeezing_action2}
\hat S^\dagger(\boldsymbol{r})\,\hat{\mathbf a}\,\hat S(\boldsymbol{r})
=
\cosh{(\boldsymbol{r})}\hat{\mathbf a}
+
\sinh{(\boldsymbol{r})}\hat{\mathbf a}^\dagger,
\end{equation}
and the corresponding nullifier equation can be expressed as
\begin{equation} \label{eq:nullifier_squeezing3}
(\hat{\mathbf a}
-
\tanh{(\boldsymbol{r})}\hat{\mathbf a}^\dagger)|\psi\rangle = 0.
\end{equation}

Thus, the nullifier equation \eqref{eq:nullifier_squeezing_displaced} together with the Takagi decomposition provides a clear connection to the key building blocks of GBS -- displacement (coherent states), squeezing and rotation (interferometer) \cite{Hamilton2017}. We use these nullifier equations to derive the operator parameters entering the corresponding displacement, squeezing and rotation operators, as shown in the derivations below.

\subsection*{Derivation of the operator parameters}

To find the operator parameters of the auxiliary state \eqref{eq:aux_state}, we first consider the
nullifier equation in the form of generalized Bogoliubov transformation \cite{Bagarello_Gen_Bogol_2015}
\begin{equation}\label{eq:first_nullifier}
\left(
\bar{\mathbf A}_\xi \hat{\mathbf a}
+\bar{\mathbf B}_\xi \hat{\mathbf a}^\dagger
+\bar{\mathbf u}_\xi
\right)|\psi_\xi^{(0)}\rangle
=
0,
\end{equation}
where the Bogoliubov matrices 
\begin{equation}
\bar{\mathbf  A}_\xi
=
\frac12
\left(
e^{-2\xi\mathbf z^T}+e^{2\xi\mathbf z}
\right),
\qquad
\bar{\mathbf  B}_\xi
=
\frac12
\left(
e^{-2\xi\mathbf z^T}-e^{2\xi\mathbf z}
\right),
\qquad
\bar{\mathbf  u}_\xi
=
-\int_0^\xi e^{-2(\xi-t)\mathbf z^T}\boldsymbol y\,dt,
\end{equation}
are found from solving a system of Heisenberg equations (see Supplementary Information).
It can be shown that for small $\xi$ $\bar{\mathbf A}_\xi$ is invertible, which leads to the following nullifier equation
\begin{equation}\label{eq:annihilation_aux}
\left(
\hat{\mathbf a}
-\bar{\mathbf Z}_\xi \hat{\mathbf a}^\dagger
+\bar{\boldsymbol\eta}_\xi
\right)|\psi_\xi^{(0)}\rangle
=
0,
\qquad
\bar{\mathbf Z}_\xi=-\bar{\mathbf A}_\xi^{-1}\bar{\mathbf B}_\xi,
\qquad
\bar{\boldsymbol\eta}_\xi=\bar{\mathbf A}_\xi^{-1}\bar{\mathbf u}_\xi.
\end{equation}
It is straightforward to show that $\bar{\mathbf Z}_\xi$ is a complex symmetric matrix, hence it must admit the Takagi factorization
\begin{equation}\label{eq:takagi_aux}
\bar{\mathbf Z}_\xi
=
\bar{\mathbf U}_\xi\,\tanh(\bar{\boldsymbol\Sigma}_\xi)\,\bar{\mathbf U}_\xi^T,
\qquad
\bar{\boldsymbol\Sigma}_\xi
=
\mathrm{diag}(\bar{\sigma}_{1,\xi},\dots,\bar{\sigma}_{M,\xi}),
\qquad
\tanh(\bar{\sigma}_{j,\xi})=\bar{s}_{j,\xi},
\end{equation}
where $\bar{\mathbf U}_\xi$ is unitary and
$\bar{s}_{j,\xi}$ are Takagi singular values of $\bar{\mathbf Z}_\xi$. Finally, the auxiliary displacement parameter is
\begin{equation}\label{eq:effective_displ_1}
\bar{\boldsymbol\gamma}_\xi
=
-(\mathbf I-\bar{\mathbf Z}_\xi\bar{\mathbf Z}_\xi^*)^{-1}
\left(
\bar{\boldsymbol\eta}_\xi+\bar{\mathbf Z}_\xi\bar{\boldsymbol\eta}_\xi^*
\right),
\end{equation}
with the full derivation available in the Supplementary Material.

All the displacement, rotation and squeezing operators in equation \eqref{eq:Aux_merging} can be combined together. This is done by considering the state $|\psi_\xi\rangle = \hat R(\mathbf{U_L})\hat S(\boldsymbol r)\hat R(\mathbf{U_R})|\psi_\xi^{(0)}\rangle$, which is the transformation of the auxiliary state $|\psi_\xi^{(0)}\rangle$ induced by part of the Doktorov operator. This state should satisfy the same type of nullifier equation as in equation \eqref{eq:first_nullifier} but with different parameters $\mathbf Z_\xi$ and $\boldsymbol\eta_\xi$: $\left(\hat{\mathbf a}-\mathbf Z_\xi \hat{\mathbf a}^\dagger+\boldsymbol\eta_\xi\right)|\psi_\xi\rangle=0$.
By conjugating it with $\hat U_0^\dagger$, one obtains $\hat U_0^\dagger\left(\hat{\mathbf a}-\mathbf Z_\xi \hat{\mathbf a}^\dagger+\boldsymbol\eta_\xi\right)\hat U_0|\psi_\xi^{(0)}\rangle=0$.
The transformation of ladder operators by $\hat U_0$ can be written in the Bogoliubov form $\hat U_0^\dagger\,\hat{\mathbf a}\,\hat U_0=\mathbf X\,\hat{\mathbf a}+\mathbf Y\,\hat{\mathbf a}^\dagger$ and $\hat U_0^\dagger\,\hat{\mathbf a}^\dagger\,\hat U_0=\mathbf Y^*\,\hat{\mathbf a}+\mathbf X^*\,\hat{\mathbf a}^\dagger$, where the corresponding Bogoliubov matrices are $\mathbf X=\mathbf{U_L}\,\cosh(\boldsymbol r)\,\mathbf{U_R},$ and $\mathbf Y=\mathbf{U_L}\,\sinh(\boldsymbol r)\,\mathbf{U_R}^*$. Hence, the transformed auxiliary state can also be written in terms of Gaussian operators as 
\begin{equation}\label{eq:psi_decomposed}
|\psi_\xi\rangle
=
\hat D(\boldsymbol\gamma_\xi)\,
\hat R(\mathbf U_\xi)\,
\hat S(\boldsymbol\Sigma_\xi)\,
|\mathbf 0\rangle.
\end{equation}
Since $\mathbf Z_\xi$ is complex symmetric, it admits the Takagi factorization
\begin{equation}\label{eq:takagi_2}
\mathbf Z_\xi
=
\mathbf U_\xi\,
\tanh(\boldsymbol\Sigma_\xi)\,
\mathbf U_\xi^T,
\qquad
\mathbf Z_\xi=\mathbf B_\xi\mathbf A_\xi^{-1}
\qquad
\boldsymbol\eta_\xi=\left(\mathbf X-\mathbf Z_\xi\mathbf Y^*\right)\bar{\boldsymbol\eta}_\xi,
\end{equation}
where $\mathbf A_\xi=\mathbf X^*+\mathbf Y^*\bar{\mathbf Z}_\xi$ and $\mathbf B_\xi=\mathbf Y+\mathbf X\bar{\mathbf Z}_\xi$. The effective displacement parameter can be found analogously in the form of equation \eqref{eq:effective_displ_1}
\begin{equation}\label{eq:gamma_tilde_def}
\boldsymbol\gamma_\xi
=
-(\mathbf I-\mathbf Z_\xi\mathbf Z_\xi^*)^{-1}
\left(
\boldsymbol\eta_\xi
+
\mathbf Z_\xi\boldsymbol\eta_\xi^*
\right).
\end{equation}
We combine the displacement operators using the Weyl composition rule $\hat D(\boldsymbol\beta)\hat D(\boldsymbol\gamma_\xi)
=e^{-i\,\mathrm{Im}(\boldsymbol\beta^\dagger\boldsymbol\gamma_\xi)}
\hat D(\boldsymbol\beta+\boldsymbol\gamma_\xi)$, and noting that the phase term disappears after taking the modulus squared, we find the total effective displacement parameter $\boldsymbol\beta_\xi=\boldsymbol\beta+\boldsymbol\gamma_\xi$. The modified operator parameters obtained here are exactly the ones that enter the auxiliary function in equation \eqref{eq:Aux_f_final_merge}.

The auxiliary scalar prefactor $C_\xi$ can be determined using vacuum overlaps
\begin{equation} \label{eq:aux_scalar_prefactor}
C_\xi
=
\frac{\langle \mathbf 0| e^{\xi\hat G_0} |\mathbf 0\rangle}{\langle \mathbf 0|\hat D(\bar{\boldsymbol\gamma}_\xi)\hat R(\bar{\mathbf U}_\xi)\hat S(\bar{\boldsymbol\Sigma}_\xi)|\mathbf 0\rangle}
=
\frac{v_\xi}{g_\xi},
\end{equation}
where
\begin{equation}\label{eq:c_xi_final}
v_\xi
=
\exp\!\left[
\int_0^\xi
\left(
\boldsymbol y^T\bar{\mathbf\eta}_t
-\frac12\,\mathrm{Tr}(\mathbf z_s\bar{\mathbf Z}_t)
-\frac12\,\bar{\boldsymbol\eta}_t^T\mathbf z_s\bar{\boldsymbol\eta}_t
\right)\,dt
\right].
\end{equation}
and
\begin{equation}\label{eq:g_xi_final}
g_\xi
=
\det(\mathbf I-\bar{\mathbf Z}_\xi\bar{\mathbf Z}_\xi^*)^{1/4}
\exp\!\left[
\frac12\bar{\boldsymbol\gamma}^\dagger_\xi\bar{\mathbf Z}_\xi\bar{\boldsymbol\gamma}^*_\xi 
-
\frac12\bar{\boldsymbol\gamma}^\dagger_\xi\bar{\boldsymbol\gamma}_\xi
\right].
\end{equation}
The scalar prefactor $C_\xi$ arises because $e^{\xi \hat{G_0}}$ is generally not a normalized unitary Gaussian transformation. When this operator is decomposed into the standard Gaussian operators in equation \eqref{eq:aux_state}, the non-commuting linear and quadratic terms generate this additional scalar factor. The detailed derivations of the operator parameters as well as the vacuum overlap integrals can be found in the Supplementary Information.

\subsection*{Condon approximation with Duschinsky rotations}

In the Condon approximation, the expansion of the derivative coupling in equation \eqref{eq:g_HT_expand} has only the constant term $g_{fi,k}(\mathbf q_0)$. This significantly simplifies the auxiliary operator because all quadratic contributions vanish
\begin{equation}
\hat G
=
\sum_{k=1}^M y_k(\hat a_k^\dagger-\hat a_k),
\end{equation}
meaning that the auxiliary state becomes $|\psi_\xi^{(0)}\rangle=C_\xi\hat D(\bar{\boldsymbol\gamma}_\xi)|\mathbf 0\rangle$, with the auxiliary displacement parameter reducing to $\bar{\boldsymbol\gamma_\xi}=\xi \boldsymbol{y}$. Hence, the corresponding auxiliary displacement operator and the scalar prefactor can be expressed as
\begin{equation}\label{eq:D_multimode_def}
\hat D(\xi \boldsymbol{y})
=
\exp\!\left[\left((\xi \boldsymbol{y})^T\hat{\mathbf a}^\dagger-(\xi \boldsymbol{y})^\dagger\hat{\mathbf a}\right)\right],
\qquad
C_\xi
=
\exp\!\left[
\frac12\left(
|\xi|^2-\xi^2
\right)\boldsymbol y^T\boldsymbol y
\right].
\end{equation}

The auxiliary displacement operator can be commuted with the Doktorov part and after merging it with $\hat D(\boldsymbol\beta)$, one obtains
\begin{equation}\label{eq:Disp_Condon}
\boldsymbol\beta_\xi
=
\boldsymbol\beta+\mathbf X(\xi\boldsymbol y)+\mathbf Y(\xi\boldsymbol y)^*,
\end{equation}
where the Bogoliubov matrices are $\mathbf X=\mathbf{U_L}\,\cosh(\boldsymbol r)\,\mathbf{U_R},$ and $\mathbf Y=\mathbf{U_L}\,\sinh(\boldsymbol r)\,\mathbf{U_R}^*$.
Therefore, in the Condon approximation, the auxiliary function reduces to
\begin{equation}\label{eq:Aux_f_condon}
f_m(\xi)
=
|C_\xi|^2
\left|
\left\langle \boldsymbol m\left|
\hat D(\boldsymbol\beta_\xi)\,
\hat R(\mathbf{U_L})\hat S(\boldsymbol r)
\right|\mathbf 0\right\rangle
\right|^2,
\end{equation}
We see that within the Condon approximation the NAC operator only modifies the effective displacement. However, the inclusion of Duschinsky mode-mixing and squeezing still requires the full set of GBS building blocks. Consequently, even this approximation remains computationally challenging, as sampling is performed from a nontrivial probability distribution governed by the loop Hafnian function \cite{Quesada2019}.

\subsection*{Linear coupling approximation}

In the simplest linear coupling approximation, the Duschinsky rotations and squeezing effects are neglected, which further simplifies the auxiliary function to
\begin{equation}\label{eq:Aux_f_linear}
f_m(\xi)
=
|C_\xi|^2
\left|
\left\langle \boldsymbol m\left|
\hat D(\boldsymbol\beta_\xi)\,
\right|\mathbf 0\right\rangle
\right|^2.
\end{equation}
In this case the Bogoliubov matrices reduce to $\mathbf{X}=\mathbf{I}$ and $\mathbf{Y}=0$, meaning that the effective displacement becomes $\boldsymbol\beta_\xi=\boldsymbol\beta+\xi\boldsymbol y$. Notably, in the linear coupling model, the squeezing and rotation operators are not required, resulting in sampling from a simple Poisson distribution \cite{EickmannVibronic2025}. The sampling task under linear coupling approximation does not pose computational challenge unlike the Condon and HT approximations with Duschinsky mode-mixing.

The so-called parallel approximation, which retains displacement and squeezing while neglecting Duschinsky rotations, provides an additional level of approximation \cite{EickmannVibronic2025}. However, we do not consider it in this work because squeezing alone has a negligible effect for DBT.

\subsection*{Herzberg-Teller approximation without Duschinsky rotations}

For the sampling simulations, the HT approximation without Duschinsky mode-mixing was obtained by setting the Duschinsky matrix to identity $\mathbf{U} = \mathbf{I}$ in equation \eqref{eq:Superposition}. Nevertheless, Valiev \textit{et al.} showed that for the HT approximation without Duschinsky rotations the formula for $k_\mathrm{IC}$ can be factorized into accepting and promoting mode parts \cite{ValievDBT}. In this case, the sampling approach would have to be applied to the accepting modes only, which are represented as simple Poisson distributions. The promoting mode part is calculated using quantum chemistry methods \cite{ValievDBT}. Therefore, we highlight that the pathway-resolved analysis in the case of HT approximation \textit{without Duschinsky rotations} is not a computationally difficult task. However, this factorized formula \textit{cannot be applied} to the case when Duschinsky rotations are included \cite{ValievDBT}, which demands the use of GBS-based methods for performing the pathway-resolved analysis.

\subsection*{Dynamic programming method with Duschinsky rotations} 

The Duschinsky rotations pose a major computational challenge for the DP method \cite{ValievDBT}. In cases when Duschinsky rotations are assumed to make small contribution, however, perturbation theory can be employed to estimate the impact of this effect \cite{Valiev2020}. The results for DP with Duschinsky rotations are presented in Table \eqref{tab:dbt_comparison}. It is still an opened question whether the Duschinsky mode-mixing effect can be considered small for DBT. This could be the root-cause which explains the difference in $k_\mathrm{IC}$ values obtained with GBS and DP methods when Duschinsky rotations are included (see Table \eqref{tab:dbt_comparison}). Based on the results presented in the main text, the effect of Duschinsky rotations for DBT is moderate but still important, since it increases the $k_\mathrm{IC}$ values by a factor of five compared to the no-Duschinsky model.

\subsection*{Details on sampling simulations}

The sampling simulations were performed for four different approximations -- Condon approximation without Duschinsky rotations (linear coupling model), Condon approximation with Duschinsky rotations, HT approximation with and without Duschinsky rotations. In all cases the final $k_\mathrm{IC}$ values were obtained by sampling each term in equation \eqref{eq:Superposition} 1 billion times. The corresponding auxiliary functions for these approximations are found in equations \eqref{eq:Aux_f_linear}, \eqref{eq:Aux_f_condon}, \eqref{eq:Aux_f_final_merge}. 

To perform the sampling simulations, it is important to determine the optimal value for the parameter $\tau$, which is described in the section below. For sampling simulations under linear coupling approximation, we used a standard desktop computer, since sampling from the Poisson distribution is a trivial task. The Walrus library \cite{Gupt2019} is used for sampling simulations beyond the linear coupling model. These simulations require significantly more computational resources, therefore we carried them out on a high-performance computing cluster.

To optimized the computational speed, the sampling simulations were performed in parallel on the computing cluster (Otus system PC2). The total number of 100 million samples was divided into independent sets of 40000 samples, each initialized with a distinct pseudo-random seed and distributed across 192 processors (2x AMD Turin 9655, up to 4.5 GHz) of a single computing node of the computing cluster. Each processor generated photon-number patterns for its assigned set and then the resulting statistics were merged together. This procedure was repeated in sequence for all terms in equation \eqref{eq:Superposition}. We ran 10 such sampling simulations on 10 computing nodes simultaneously and combined their statistics to get the final 1 billion samples per term in equation \eqref{eq:Superposition}. The sampling data is obtained for the photon-number cutoff of 10 per mode and limited to 240 total number of photons across all 168 modes. All sampling outcomes that exceed these limits are discarded and sampled again. With all the settings mentioned above, the total computational time resulted in approximately 10 hours provided that the simulations in all 10 computing nodes started simultaneously.

\subsection*{Determining optimal value for parameter $\tau$}

\begin{figure}[!ht] 
    \centering
    \includegraphics[width = \textwidth]{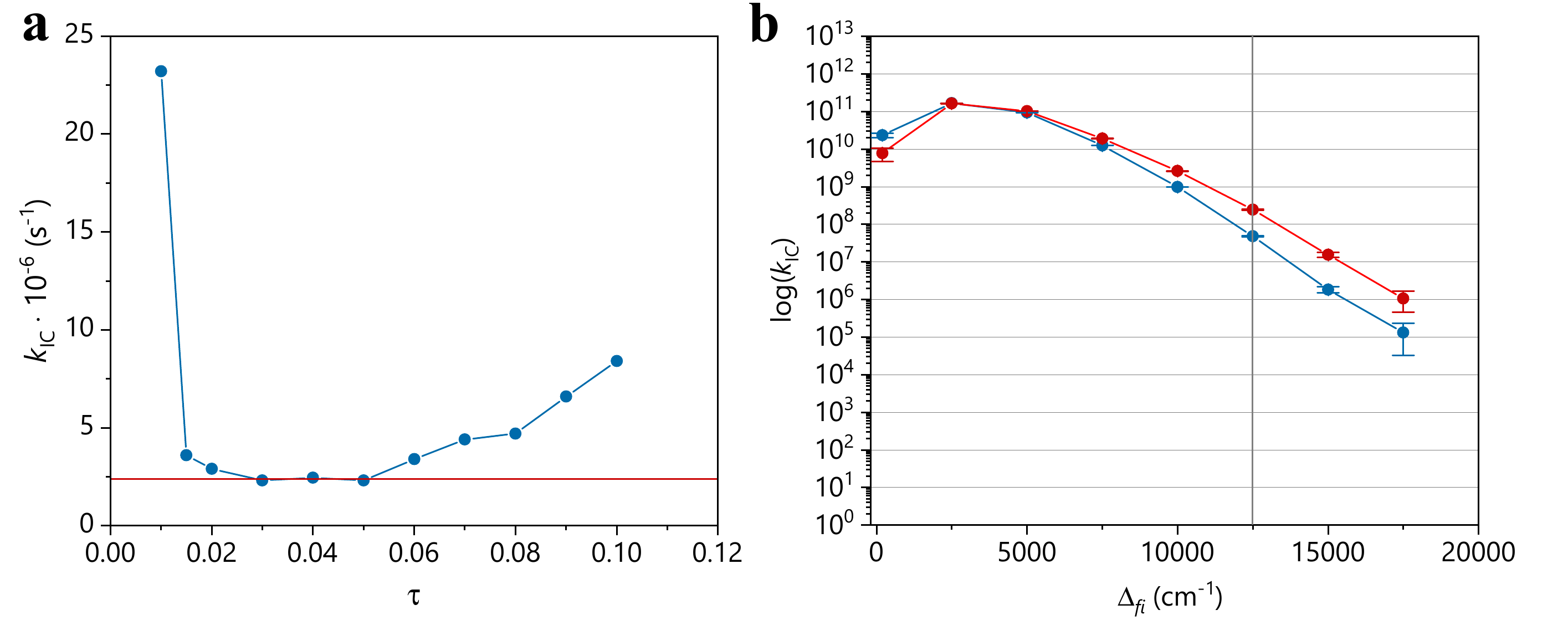}
    \caption{
    (a) $k_\mathrm{IC}$ values for DBT as a function of $\tau$ within linear coupling approximation for $\Delta_{fi}$=12500 cm$^{-1}$. The red line shows the reference data from quantum chemistry $k_\mathrm{IC} = 2.4\cdot 10^{6}$ s$^{-1}$ obtained using dynamic programming method under Condon approximation \cite{ValievDBT}. (b) $k_\mathrm{IC}$ as a function of energy gap $\Delta_{fi}$ in two scenarios: HT approximation without Duschinsky rotations (blue dots) and HT approximation with Duschinsky rotations (red dots). The grey vertical line represents the real energy gap of 12500 cm$^{-1}$ for the DBT molecule \cite{MishraDBT2025}.}
    \label{Fig4}
\end{figure}

To optimize the numerical stability of our GBS framework, it is important to determine an optimal value of the parameter $\tau$. This can be achieved efficiently by running multiple sampling simulations with different values of $\tau$ using the linear coupling model, the results of which are shown in Figure \ref{Fig4} (a). Each point on the graph corresponds to a sampling simulation according to equations \eqref{eq:Superposition} and \eqref{eq:Aux_f_linear} with 100 million samples per term. The graph indicates that the obtained $k_\mathrm{IC}$ values reach a plateau within the range $0.03 \leq \tau \leq 0.05$, meaning that the rate constant calculation is numerically stable in this region. We therefore used $\tau=0.03$ in all sampling simulations presented in the main text. Notably, the average rate constant value $2.3\cdot 10^{6}$ s$^{-1}$ in this region agrees very well with results from the DP method $2.4\cdot 10^{6}$ s$^{-1}$, indicated by the red horizontal line.

\subsection*{Calculation of $k_\mathrm{IC}$ values}

To calculate $k_\mathrm{IC}$, we use a flat-window approximation of Plotnikov's formula \cite{Valiev2020}
\begin{equation}\label{eq:k_flat_window}
k_{\mathrm{IC}}
=
\frac{4 p^2}{\Gamma_{fm}}\sum_{\boldsymbol{m}\in Q_{\Delta E}}\left| \langle \boldsymbol{m} | \hat{U}_{Dok}\,\hat{G} | \boldsymbol{0} \rangle\right|^2,
\quad
Q_{\Delta E} =\{\boldsymbol{m}:|\Delta_{\boldsymbol{m}}|\le \Delta E\},
\quad
\Delta_{\boldsymbol{m}}
=
\sum_{k=1}^Mm_k\omega^\prime_k - \Delta_{fi}.
\end{equation}
To allow direct comparison of our results with the $k_\mathrm{IC}$ values from the literature \cite{ValievDBT}, we used $\Delta E = 200~\mathrm{cm}^{-1}$ for the half-width of the energy window and $\Gamma_{fm} = 7.8 \cdot 10^{13}$ as the scaling prefactor. Each sampled photon-number pattern $\boldsymbol{m}=(m_1, \dots,m_k, \dots, m_M)$ corresponds to a final-state vibrational energy $\sum_{k=1}^Mm_k\omega^\prime_k$, therefore $\Delta_{\boldsymbol{m}}$ represents the detuning from the electronic energy gap $\Delta_{fi}$, which for DBT was experimentally measured to be $\sim$12500 cm$^{-1}$ \cite{MishraDBT2025}. The dimensionless NACME is assessed via sampling using equation \eqref{eq:Superposition} in combination with the auxiliary functions in equations \eqref{eq:Aux_f_final_merge}, \eqref{eq:Aux_f_condon}, \eqref{eq:Aux_f_linear} each corresponding to a different approximation. The NACME is then multiplied by a normalization factor defined as $p=\max\left(\max_{k}\left|\widetilde{y}_k\right|,\max_{j,k}\left|\widetilde{z}_{j,k}\right|\right)$ to recover the dimensional scale.

In the GBS setting, to calculate $k_\mathrm{IC}$ one needs to estimate the sum of squared NACME values $|V_{i0,f\boldsymbol m}|^{2}$ for each photon pattern $\boldsymbol{m}$ that satisfies the energy gap condition mentioned above. This can be calculated using
\begin{equation} \label{eq:NACME_pattern}
\sum_{\boldsymbol{m}\in Q_{\Delta E}}|V_{i0,f\boldsymbol m}|^{2}
\approx
\frac{1}{2\tau^{2}}\sum_{\boldsymbol{m}\in Q_{\Delta E}}
\left[
f_{\boldsymbol{m}}(i\tau)+\frac{1}{2}f_{\boldsymbol{m}}(\tau)+\frac{1}{2}f_{\boldsymbol{m}}(-\tau)-2f_{\boldsymbol{m}}(0)
\right],
\end{equation}
however, there is another mathematically equivalent form
\begin{equation} \label{eq:NACME_avg}
\sum_{\boldsymbol{m}\in Q_{\Delta E}}|V_{i0,f\boldsymbol m}|^{2}
\approx
\frac{1}{2\tau^{2}}
\left[
\sum_{\boldsymbol{m}\in Q_{\Delta E}}f_{\boldsymbol{m}}(i\tau)+\sum_{\boldsymbol{m}\in Q_{\Delta E}}\frac{1}{2}f_{\boldsymbol{m}}(\tau)+\sum_{\boldsymbol{m}\in Q_{\Delta E}}\frac{1}{2}f_{\boldsymbol{m}}(-\tau)-\sum_{\boldsymbol{m}\in Q_{\Delta E}}2f_{\boldsymbol{m}}(0)
\right].
\end{equation}
Importantly, if these two forms are used to calculate $k_\mathrm{IC}$ according to equation \eqref{eq:k_flat_window}, it is possible that the resulting $k_\mathrm{IC}$ values will not match. This can happen because equation \eqref{eq:NACME_pattern} is more vulnerable to sampling fluctuations. For very rare photon patterns that happen to appear only in $f_m(0)$ term, the corresponding squared NACME values calculated via \eqref{eq:NACME_pattern} would be assigned a negative value. Such results are unphysical and need to be discarded. However, this introduces an upward bias in the final $k_\mathrm{IC}$ values. Equation \eqref{eq:NACME_avg} is not as sensitive to undersampling because all the photon-pattern contributions are first summed up in each term, and only after that the difference is evaluated. For the infinite number of samples, the pattern-resolved and averaged forms converge to the same value.

\subsection*{Duschinsky rotations in molecules with large energy gaps}

To further assess the role of Duschinsky rotations, we performed an artificial scan of the energy gap and recalculated the corresponding $k_\mathrm{IC}$ values. For DBT, the relevant energy gap is comparatively modest ($12500~\mathrm{cm}^{-1}$), which explains why Duschinsky rotations provide only a moderate contribution. However, this effect is expected to become increasingly more important for molecules with larger energy gaps. This trend is exemplified in Figure \ref{Fig4}(b): as the energy gap is increased towards $20000~\mathrm{cm}^{-1}$, the inclusion of Duschinsky rotations enhances $k_\mathrm{IC}$ by more than an order of magnitude relative to the no-Duschinsky model \cite{ValievDBT}. This confirms that Duschinsky rotations become progressively more important as higher-energy vibronic configurations are involved in the IC process \cite{ValievDBT}.

\end{document}


\maketitle

\section*{1. Quantum optics operators and nullifier equations}

The multimode displacement, rotation and squeezing  operators are defined as \cite{MaRhodes1990, Jnane2021}
\begin{equation} \label{eq:Disp_op_define}
\hat D(\boldsymbol{\gamma})
=
\exp\!\left(
\boldsymbol{\gamma}^T \hat{\mathbf a}^\dagger
-
\boldsymbol{\gamma}^\dagger \hat{\mathbf a}
\right),
\end{equation}
\begin{equation} \label{eq:Rotation_op_define}
\hat R(\mathbf U)
=
\exp\!\left[
\hat{\mathbf a}^{\dagger T}\bigl(\ln \mathbf U\bigr)\hat{\mathbf a}
\right],
\end{equation}
\begin{equation} \label{eq:MMSqueezing_op_define}
\hat S(\boldsymbol{\Xi})
=
\exp\!\left[
\frac12
\left(
\hat{\mathbf a}^{\dagger T}\boldsymbol{\Xi}\hat{\mathbf a}^\dagger
-
\hat{\mathbf a}^T\boldsymbol{\Xi}^*\hat{\mathbf a}
\right)
\right],
\end{equation}
where $\mathbf{a}=(a_1, \dots, a_M)^T$ and $\mathbf{a^\dagger}=(a_1^\dagger, \dots, a_M^\dagger)^T$ are the multi-mode ladder operators \cite{MaRhodes1990}. The action of the displacement, rotation and squeezing operators on the ladder operators is given by \cite{MaRhodes1990}
\begin{equation} \label{eq:Disp_action}
\hat D^\dagger(\boldsymbol{\gamma})\,\hat{\mathbf a}\,\hat D(\boldsymbol{\gamma})
=
\hat{\mathbf a}+\boldsymbol{\gamma},
\end{equation}
\begin{equation} \label{eq:Rotation_action}
\hat R^\dagger(\mathbf U)\,\hat{\mathbf a}\,\hat R(\mathbf U)
=
\mathbf U\,\hat{\mathbf a},
\end{equation}
\begin{equation} \label{eq:Squeezing_action}
\hat S^\dagger(\boldsymbol{\Xi})\,\hat{\mathbf a}\,\hat S(\boldsymbol{\Xi})
=
\mathbf{A}\hat{\mathbf a}
+
\mathbf{B}\hat{\mathbf a}^\dagger.
\end{equation}
Here $\boldsymbol{\gamma}$ is a complex vector, $\mathbf U$ is a unitary, $\boldsymbol{\Xi}$ is a complex symmetric matrix, and matrices $\mathbf{A}$ and $\mathbf{B}$ are the Bogoliubov matrices. A common way to write the Bogoliubov matrices in this case is using the polar decomposition \cite{MaRhodes1990}
\begin{equation} \label{eq:Squeezing_action_MaRhodes}
\hat S^\dagger(\boldsymbol{\Xi})\,\hat{\mathbf a}\,\hat S(\boldsymbol{\Xi})
=
\cosh{(\boldsymbol{r}_{\mathrm{pol}})}\hat{\mathbf a}
+
\sinh{(\boldsymbol{r}_{\mathrm{pol}})}e^{i\boldsymbol \theta}\hat{\mathbf a}^\dagger,
\end{equation}
where $\boldsymbol{\Xi} = \boldsymbol{r}_{\mathrm{pol}}e^{i\boldsymbol{\theta}}$ with $\boldsymbol{r}_{\mathrm{pol}}$ being the positive semi-definite matrix representing the squeezing magnitudes and $e^{i\boldsymbol \theta}$ encodes squeezing phases \cite{MaRhodes1990}.

However, in this work we use an alternative representation of equation \eqref{eq:Squeezing_action_MaRhodes}. Since $\boldsymbol{\Xi}$ is complex symmetric matrix, it must admit the Takagi decomposition $\boldsymbol{\Xi}=\mathbf{W}\boldsymbol{r}\mathbf{W}^T$ \cite{CariolaroTAKAGI}, where $\boldsymbol{r} = diag(r_1, \dots ,r_M)$ is the diagonal matrix of single-mode squeezing strengths (real non-negative values) and $\mathbf{W}$ is a unitary, which represents the basis change that leads to the multimode correlation of the squeezed states. Inserting identity matrix $\mathbf{W}^{\dagger}\mathbf{W}$ into the Takagi decomposition $\boldsymbol{\Xi}=\mathbf{W}\boldsymbol{r}\mathbf{W}^{\dagger}\mathbf{W}\mathbf{W}^T$, the correspondence between the polar and Takagi decompositions becomes clear: $\boldsymbol{r}_{\mathrm{pol}}=\mathbf{W}\boldsymbol{r}\mathbf{W}^{\dagger}$, $e^{i\theta} = \mathbf{W}\mathbf{W}^T$. This leads to the Takagi decomposed form of the Bogoliubov transformation
\begin{equation} \label{eq:Squeezing_action_Takagi}
\hat S^\dagger(\boldsymbol{\Xi})\,\hat{\mathbf a}\,\hat S(\boldsymbol{\Xi})
=
\mathbf{W}\cosh{(\boldsymbol{r})}\mathbf{W}^{\dagger}\hat{\mathbf a}
+
\mathbf{W}\sinh{(\boldsymbol{r})}\mathbf{W}^T\hat{\mathbf a}^\dagger.
\end{equation}

Now we introduce the nullifier equation \cite{LasserNULLIFIERS2018,FunaiNULLIFIERS2025} for the general multimode squeezed state. Starting from the vacuum condition \(\hat{\mathbf a}|\mathbf 0\rangle=0\), one has
\begin{equation} \label{eq:nullifier_squeezing}
0
=
\hat S(\boldsymbol{\Xi})\hat{\mathbf a}|\mathbf 0\rangle
=
\left(\hat S(\boldsymbol{\Xi})\hat{\mathbf a}\hat S^\dagger(\boldsymbol{\Xi})\right)\hat S(\boldsymbol{\Xi})|\mathbf 0\rangle
=
\left(\hat S(\boldsymbol{\Xi})\hat{\mathbf a}\hat S^\dagger(\boldsymbol{\Xi})\right)|\psi\rangle,
\end{equation}
where $|\psi\rangle$ is the squeezed vacuum state. Using $S^\dagger(\boldsymbol{\Xi})=S(\boldsymbol{-\Xi})$ and the fact that Takagi decomposition of the negative matrix is $-\boldsymbol{\Xi}=(i\mathbf{W})\boldsymbol{r}(i\mathbf{W}^T)$, it is straightforward to obtain the action entering equation \eqref{eq:nullifier_squeezing}:
\begin{equation} \label{eq:Squeezing_action_Takagi_reversed}
\hat S(\boldsymbol{\Xi})\hat{\mathbf a}\,\hat S^\dagger(\boldsymbol{\Xi})
=
\mathbf{W}\cosh{(\boldsymbol{r})}\mathbf{W}^{\dagger}\hat{\mathbf a}
-
\mathbf{W}\sinh{(\boldsymbol{r})}\mathbf{W}^T\hat{\mathbf a}^\dagger.
\end{equation}
Hence, we obtain the corresponding nullifier equation
\begin{equation} \label{eq:nullifier_squeezing2}
(\mathbf{W}\cosh{(\boldsymbol{r})}\mathbf{W}^{\dagger}\hat{\mathbf a}
-
\mathbf{W}\sinh{(\boldsymbol{r})}\mathbf{W}^T\hat{\mathbf a}^\dagger)|\psi\rangle = 0.
\end{equation}
Since $\mathbf{W}\cosh{(\boldsymbol{r})}\mathbf{W}^{\dagger}$ is invertible, which follows from $\cosh{(r_j)\ge 1}$, we can write the nullifier equation in the following form:
\begin{equation} \label{eq:nullifier_squeezing4}
\left(
\hat{\mathbf a}
-
\mathbf Z\,\hat{\mathbf a}^\dagger
\right)|\psi\rangle=0,
\qquad
\mathbf Z=\mathbf W\tanh(\boldsymbol r)\mathbf W^T.
\end{equation}

For the operator $\hat S(\boldsymbol{\Xi})$ the Bloch-Messiah decomposition can be written as
\begin{equation}\label{eq:multimode_squeezing_decomposition}
\hat S(\boldsymbol{\Xi}) = \hat R(\mathbf{W})\hat S(\boldsymbol{r}) \hat R^\dagger(\mathbf{W}),
\end{equation}
which using $\hat R^\dagger(\mathbf{W})|\mathbf 0\rangle = |\mathbf 0\rangle$ leads to
\begin{equation}\label{eq:multimode_squeezing_decomposition_2}
|\psi\rangle = \hat S(\boldsymbol{\Xi})|\mathbf 0\rangle = \hat R(\mathbf{W})\hat S(\boldsymbol{r})|\mathbf 0\rangle.
\end{equation}

For the general displaced squeezed multimode state we can write
\begin{equation}\label{eq:multimode_squeezing_decomposition_displaced}
|\psi_{\gamma}\rangle=\hat D(\boldsymbol{\gamma})|\psi\rangle = \hat D(\boldsymbol{\gamma})\hat R(\mathbf{W})\hat S(\boldsymbol{r})|\mathbf 0\rangle.
\end{equation}
To obtain the corresponding nullifier equation in this case, we start from equation \eqref{eq:nullifier_squeezing4} and transform it using the displacement operator in equation \eqref{eq:Disp_op_define}
\begin{equation} \label{eq:nullifier_squeezing_displaced_1}
D(\boldsymbol{\gamma})\left(
\hat{\mathbf a}
-
\mathbf Z\,\hat{\mathbf a}^\dagger
\right)D^{\dagger}(\boldsymbol{\gamma})|\psi_{\gamma}\rangle=0.
\end{equation}
From equation \eqref{eq:Disp_action} it follows that $\hat D(\boldsymbol{\gamma})\,\hat{\mathbf a}\,\hat D^\dagger(\boldsymbol{\gamma}) = \hat{\mathbf{a}}-\gamma$ and $\hat D(\boldsymbol{\gamma})\,\hat{\mathbf a}^\dagger\,\hat D^\dagger(\boldsymbol{\gamma}) = \hat{\mathbf{a}}^\dagger-\gamma^*$, which leads to
\begin{equation} \label{eq:nullifier_squeezing_displaced}
\left(
\hat{\mathbf a}
-
\mathbf Z\,\hat{\mathbf a}^\dagger
+
\boldsymbol \eta
\right)|\psi_{\gamma}\rangle=0,
\end{equation}
where we introduced $\boldsymbol{\eta}=\mathbf Z\gamma^*-\gamma$.

Additionally, we note that in the special case of independent single-mode squeezers when $\boldsymbol{\Xi}=\boldsymbol{r}$, operator $\hat S(\boldsymbol{\Xi})$ becomes a product of independent single-mode squeezers, whose action on the ladder operator reduces to \cite{MaRhodes1990}
\begin{equation} \label{eq:Squeezing_action2}
\hat S^\dagger(\boldsymbol{r})\,\hat{\mathbf a}\,\hat S(\boldsymbol{r})
=
\cosh{(\boldsymbol{r})}\hat{\mathbf a}
+
\sinh{(\boldsymbol{r})}\hat{\mathbf a}^\dagger,
\end{equation}
and the corresponding nullifier equation can be expressed as
\begin{equation} \label{eq:nullifier_squeezing3}
(\hat{\mathbf a}
-
\tanh{(\boldsymbol{r})}\hat{\mathbf a}^\dagger)|\psi\rangle = 0.
\end{equation}

Thus, the nullifier equation \eqref{eq:nullifier_squeezing_displaced} together with the Takagi decomposition provides a clear connection to the key building blocks of GBS -- displacement (coherent states), squeezing and rotation (interferometer) \cite{Hamilton2017}. We use these nullifier equations to derive the operator parameters entering the corresponding displacement, squeezing and rotation operators, as shown in the derivations below.

\section*{2. Derivation of the auxiliary operator parameters}

\subsection*{2a. Definition of the auxiliary operator}

The general form of the auxiliary operator writes
\begin{equation}\label{eq:Ggen_canonical_full}
\hat G
=
\mathrm{Tr}(\mathbf{z})
+
\left(\boldsymbol{y}^T\hat{\mathbf a}^\dagger-\boldsymbol{y}^T\hat{\mathbf a}\right)
+
\frac{1}{2}\left(
\hat{\mathbf a}^{\dagger T}\mathbf{z}_s\hat{\mathbf a}^\dagger
-\hat{\mathbf a}^{T}\mathbf{z}_s\hat{\mathbf a}\right)
-\hat{\mathbf a}^{\dagger T}\mathbf{z}_a\hat{\mathbf a}
=
\mathrm{Tr}(\mathbf{z})+ \hat G_0,
\end{equation}
where $\boldsymbol{y}=(y_1, \dots, y_M)^T$ is a dimensionless vector and $\mathbf{z}=(z_{j,k})$ is a dimensionless matrix. The matrix $\mathbf{z}$ is decomposed into its symmetric $\mathbf{z}_s=\mathbf{z}+\mathbf{z}^T$ and anti-symmetric $\mathbf{z}_a=\mathbf{z}-\mathbf{z}^T$ counterparts. We define $y_k = \Tilde{y}_k/p$ and $z_{j,k} = \Tilde{z}_{j,k}/p$ where $p$ is the normalization factor having the same dimensions as $\Tilde{y}_k$ and $\Tilde{z}_{j,k}$.
The quantities that have physical dimensions are $\Tilde{y}_k \in \mathbb{R}$ and $\Tilde{z}_{jk} \in \mathbb{C}$ defined as
\begin{equation}\label{eq:yz_defs}
\Tilde{y}_k = \hbar^2\sqrt{\frac{\omega_k}{2\hbar}} g_{fi,k}(\mathbf q_0),
\qquad
\Tilde{z}_{jk}=\frac{\hbar^2}{2}\left.\sqrt{\frac{\omega_k}{\omega_j}}\frac{\partial g_{fi,k}(\mathbf q)}{\partial q_j}\right|_{\mathbf q_0}.
\end{equation}
where $\omega_k$ and $\omega_j$ are the final frequencies for modes $k$ and $j$.

To determine the auxiliary operator parameters, we consider the action of the auxiliary operator on the vacuum state:
\begin{equation}\label{eq:auxiliary_Gaussian_state_general}
|\psi_\xi\rangle
=
e^{\xi \hat G}|\mathbf 0\rangle
=
e^{\xi\,\mathrm{Tr}(\mathbf{z})}\,e^{\xi \hat G_0}|\mathbf 0\rangle.
\end{equation}
Hence, the nontrivial part is to study the state
\begin{equation} \label{eq:psy_0_state}
|\psi_\xi^{(0)}\rangle
=
e^{\xi \hat G_0}|\mathbf 0\rangle.
\end{equation}
In the next sections, it will be shown how this state can be decomposed into the building blocks of Gaussian boson sampling -- displacement, squeezing and rotation operators.

\subsection*{2b. Commutation relations of the transformed mode operators}

To characterize the state in equation \eqref{eq:psy_0_state}, we study how the operator \(\hat G_0\) transforms the bosonic mode operators. We define the transformed annihilation and creation operators as
\begin{equation} \label{transformed_operators}
\hat{\mathbf b}_\xi
=
e^{\xi\hat G_0}\hat{\mathbf a}\,e^{-\xi\hat G_0},
\qquad
\hat{\mathbf c}_\xi
=
e^{\xi\hat G_0}\hat{\mathbf a}^\dagger e^{-\xi\hat G_0}.
\end{equation}
With these definitions, the nullifier equation of the transformed state follows immediately $\hat{\mathbf b}_\xi |\psi_\xi^{(0)}\rangle = 0$. These transformed operators preserve the canonical commutation relations, even though they need not be Hermitian adjoints of each other. To show this, we define
\begin{equation}
\hat T_\xi=e^{\xi\hat G_0},
\qquad
\hat b_{\xi,i}=\hat T_\xi\hat a_i\hat T_\xi^{-1},
\qquad
\hat c_{\xi,j}=\hat T_\xi\hat a_j^\dagger\hat T_\xi^{-1}.
\end{equation}
We can write their corresponding commutation relation as
\begin{align}
[\hat b_{\xi,i},\hat c_{\xi,j}]
&=
\hat T_\xi\hat a_i\hat T_\xi^{-1}
\hat T_\xi\hat a_j^\dagger\hat T_\xi^{-1}
-
\hat T_\xi\hat a_j^\dagger\hat T_\xi^{-1}
\hat T_\xi\hat a_i\hat T_\xi^{-1}
\\
&=
\hat T_\xi
\left(\hat a_i\hat a_j^\dagger
-\hat a_j^\dagger\hat a_i\right)
\hat T_\xi^{-1}
\\
&=
\hat T_\xi[\hat a_i,\hat a_j^\dagger]\hat T_\xi^{-1}
=
\delta_{ij}\hat I.
\label{eq:comm_b_c}
\end{align}
Similarly one can show
\begin{align}
[\hat b_{\xi,i},\hat b_{\xi,j}]
&=
\hat T_\xi[\hat a_i,\hat a_j]\hat T_\xi^{-1}
=0,
\label{eq:comm_b_b}
\\
[\hat c_{\xi,i},\hat c_{\xi,j}]
&=
\hat T_\xi[\hat a_i^\dagger,\hat a_j^\dagger]
\hat T_\xi^{-1}
=0.
\label{eq:comm_c_c}
\end{align}
To describe these transformed operators, we use the generalized Bogoliubov transformation \cite{Bagarello_Gen_Bogol_2015}. We first need to determine the dependence of $\hat{\mathbf b}_\xi$ and $\hat{\mathbf c}_\xi$ on the auxiliary parameter $\xi$, which can be obtained by differentiation. For a generic operator $\hat O$, one can define
\begin{equation}
\hat O(\xi)
=
e^{\xi\hat G_0}\hat O\,e^{-\xi\hat G_0},
\end{equation}
and applying the product rule yields
\begin{align}
\frac{d}{d\xi}\hat O(\xi)
&=
\frac{d}{d\xi}\!\left(e^{\xi\hat G_0}\right)\hat O\,e^{-\xi\hat G_0}
+
e^{\xi\hat G_0}\hat O\,\frac{d}{d\xi}\!\left(e^{-\xi\hat G_0}\right) \\
&=
\hat G_0 e^{\xi\hat G_0}\hat O e^{-\xi\hat G_0}
-
e^{\xi\hat G_0}\hat O e^{-\xi\hat G_0}\hat G_0 \\
&=
e^{\xi\hat G_0}\bigl(\hat G_0\hat O-\hat O\hat G_0\bigr)e^{-\xi\hat G_0} \\
&=
e^{\xi\hat G_0}\bigl([\hat G_0,\hat O]\bigr)e^{-\xi\hat G_0}.
\end{align}
Thus, for \(\hat O=\hat{\mathbf a}\) and \(\hat O=\hat{\mathbf a}^\dagger\), one obtains
\begin{equation} \label{eq:Diff_equations}
\frac{d}{d\xi}\hat{\mathbf b}_\xi
=
e^{\xi\hat G_0}[\hat G_0,\hat{\mathbf a}]e^{-\xi\hat G_0},
\qquad
\frac{d}{d\xi}\hat{\mathbf c}_\xi
=
e^{\xi\hat G_0}[\hat G_0,\hat{\mathbf a}^\dagger]e^{-\xi\hat G_0},
\end{equation}
which are the Heisenberg equations generated by \(\hat G_0\), with \(\xi\) playing the role of an auxiliary evolution parameter.

The required commutators can be derived using the canonical commutation relations
\begin{equation}
[\hat a_j,\hat a_k^\dagger]=\delta_{jk},
\qquad
[\hat a_j,\hat a_k]=0,
\qquad
[\hat a_j^\dagger,\hat a_k^\dagger]=0.
\end{equation}
First, for the commutator with \(\hat a_\ell\), we treat each term in \(\hat G_0\) separately. For the linear contribution in \eqref{eq:Ggen_canonical_full} one has
\begin{equation}
\left[\sum_j y_j\hat a_j^\dagger-\sum_j y_j\hat a_j,\hat a_\ell\right]
=
\sum_j y_j[\hat a_j^\dagger,\hat a_\ell]
-
\sum_j y_j[\hat a_j,\hat a_\ell]
=
-\sum_j y_j\delta_{j\ell}
=
-y_\ell.
\end{equation}
For the pair-creation term, one has
\begin{align}
\left[\frac12\sum_{jk}(\boldsymbol{z}_s)_{jk}\hat a_j^\dagger\hat a_k^\dagger,\hat a_\ell\right]
&=
\frac12\sum_{jk}(\boldsymbol{z}_s)_{jk}
\left(
\hat a_j^\dagger[\hat a_k^\dagger,\hat a_\ell]
+
[\hat a_j^\dagger,\hat a_\ell]\hat a_k^\dagger
\right) \\
&=
-\frac12\sum_{jk}(\boldsymbol{z}_s)_{jk}
\left(
\delta_{k\ell}\hat a_j^\dagger+\delta_{j\ell}\hat a_k^\dagger
\right).
\end{align}
Since $\mathbf{z}_s$ is symmetric, the last two terms are identical, and therefore
\begin{equation}
\left[\frac12\hat{\mathbf a}^{\dagger T}\mathbf{z}_s\hat{\mathbf a}^\dagger,\hat a_\ell\right]
=
-\sum_j(\boldsymbol{z}_s)_{\ell j}\hat a_j^\dagger
=
-(\boldsymbol{z}_s\hat{\mathbf a}^\dagger)_\ell.
\end{equation}
The pair-annihilation term gives no contribution, because annihilation operators mutually commute:
\begin{equation}
\left[-\frac12\hat{\mathbf a}^{T}\mathbf{z}_s\hat{\mathbf a},\hat a_\ell\right]
=
0.
\end{equation}
Finally, for the number-conserving quadratic term,
\begin{align}
\left[-\sum_{jk}(\boldsymbol{z}_a)_{jk}\hat a_j^\dagger\hat a_k,\hat a_\ell\right]
&=
-\sum_{jk}(\boldsymbol{z}_a)_{jk}
\left(
\hat a_j^\dagger[\hat a_k,\hat a_\ell]
+
[\hat a_j^\dagger,\hat a_\ell]\hat a_k
\right) \\
&=
\sum_{jk}(\boldsymbol{z}_a)_{jk}\delta_{j\ell}\hat a_k
=
\sum_k(\boldsymbol{z}_a)_{\ell k}\hat a_k
=
(\boldsymbol{z}_a\hat{\mathbf a})_\ell.
\end{align}
Collecting all contributions, we obtain
\begin{equation}
[\hat G_0,\hat a_\ell]
=
(\boldsymbol{z}_a\hat{\mathbf a})_\ell
-
(\boldsymbol{z}_s\hat{\mathbf a}^\dagger)_\ell
-
y_\ell,
\end{equation}
or, in vector form,
\begin{equation} \label{eq:comm_G0_a}
[\hat G_0,\hat{\mathbf a}]
=
\mathbf{z}_a\,\hat{\mathbf a}
-\mathbf{z}_s\,\hat{\mathbf a}^\dagger
-\boldsymbol{y}.
\end{equation}

The commutator with \(\hat a_\ell^\dagger\) is derived similarly. The linear term gives
\begin{equation}
\left[\sum_j y_j\hat a_j^\dagger-\sum_j y_j\hat a_j,\hat a_\ell^\dagger\right]
=
-\sum_j y_j[\hat a_j,\hat a_\ell^\dagger]
=
-y_\ell.
\end{equation}
The pair-creation term vanishes since creation operators commute:
\begin{equation}
\left[\frac12\hat{\mathbf a}^{\dagger T}\boldsymbol{z}_s\hat{\mathbf a}^\dagger,\hat a_\ell^\dagger\right]
=
0.
\end{equation}
For the pair-annihilation term,
\begin{align}
\left[-\frac12\sum_{jk}(\boldsymbol{z}_s)_{jk}\hat a_j\hat a_k,\hat a_\ell^\dagger\right]
&=
-\frac12\sum_{jk}(\boldsymbol{z}_s)_{jk}
\left(
\hat a_j[\hat a_k,\hat a_\ell^\dagger]
+
[\hat a_j,\hat a_\ell^\dagger]\hat a_k
\right) \\
&=
-\frac12\sum_{jk}(\boldsymbol{z}_s)_{jk}
\left(
\delta_{k\ell}\hat a_j+\delta_{j\ell}\hat a_k
\right).
\end{align}
Since \(\boldsymbol{z}_s\) is symmetric, this becomes
\begin{equation}
\left[-\frac12\hat{\mathbf a}^{T}\boldsymbol{z}_s\hat{\mathbf a},\hat a_\ell^\dagger\right]
=
-\sum_j(\boldsymbol{z}_s)_{\ell j}\hat a_j
=
-(\boldsymbol{z}_s\hat{\mathbf a})_\ell.
\end{equation}
Finally,
\begin{align}
\left[-\sum_{jk}(\boldsymbol{z}_a)_{jk}\hat a_j^\dagger\hat a_k,\hat a_\ell^\dagger\right]
&=
-\sum_{jk}(\boldsymbol{z}_a)_{jk}
\left(
\hat a_j^\dagger[\hat a_k,\hat a_\ell^\dagger]
+
[\hat a_j^\dagger,\hat a_\ell^\dagger]\hat a_k
\right) \\
&=
-\sum_{jk}(\boldsymbol{z}_a)_{jk}\hat a_j^\dagger\delta_{k\ell}
=
-\sum_j(\boldsymbol{z}_a)_{j\ell}\hat a_j^\dagger
=
-(\boldsymbol{z}_a^T\hat{\mathbf a}^\dagger)_\ell.
\end{align}
Hence we obtain
\begin{equation}
[\hat G_0,\hat a_\ell^\dagger]
=
-(\boldsymbol{z}_s\hat{\mathbf a})_\ell
-(\boldsymbol{z}_a^T\hat{\mathbf a}^\dagger)_\ell
-y_\ell,
\end{equation}
or in the vector form
\begin{equation}
[\hat G_0,\hat{\mathbf a}^\dagger]
=
-\mathbf{z}_s\,\hat{\mathbf a}
-\mathbf{z}_a^T\,\hat{\mathbf a}^\dagger
-\boldsymbol{y}.
\end{equation}
Since \(\boldsymbol{z}_a\) is antisymmetric, \(\boldsymbol{z}_a^T=-\boldsymbol{z}_a\), this simplifies to
\begin{equation} \label{eq:Comm_G0_a_dagger}
[\hat G_0,\hat{\mathbf a}^\dagger]
=
-\mathbf{z}_s\,\hat{\mathbf a}
+\mathbf{z}_a\,\hat{\mathbf a}^\dagger
-\boldsymbol{y}.
\end{equation}

\subsection*{2c. Coupled differential equations of transformed mode operators}

Using the commutation relations in equations \eqref{eq:comm_G0_a} and \eqref{eq:Comm_G0_a_dagger} together with equations \eqref{eq:Diff_equations}, we obtain the coupled ordinary differential equations
\begin{equation}
\left\{
\begin{aligned}
\frac{d}{d\xi}\hat{\mathbf b}_\xi
&=
\mathbf{z}_a\,\hat{\mathbf b}_\xi
-\mathbf{z}_s\,\hat{\mathbf c}_\xi
-\boldsymbol{y},
\\[1mm]
\frac{d}{d\xi}\hat{\mathbf c}_\xi
&=
-\mathbf{z}_s\,\hat{\mathbf b}_\xi
+\mathbf{z}_a\,\hat{\mathbf c}_\xi
-\boldsymbol{y}.
\end{aligned}
\right.
\end{equation}
To solve this system, we define the sum and difference operators
\begin{equation}
\hat{\mathbf p}_\xi=\hat{\mathbf b}_\xi+\hat{\mathbf c}_\xi,
\qquad
\hat{\mathbf q}_\xi=\hat{\mathbf b}_\xi-\hat{\mathbf c}_\xi.
\end{equation}
Differentiating $\hat{\mathbf p}_\xi$ and substituting the coupled equations gives
\begin{align}
\frac{d}{d\xi}\hat{\mathbf p}_\xi
&=
\frac{d}{d\xi}\hat{\mathbf b}_\xi
+
\frac{d}{d\xi}\hat{\mathbf c}_\xi
\\
&=
\mathbf{z}_a\,\hat{\mathbf b}_\xi
-\mathbf{z}_s\,\hat{\mathbf c}_\xi
-\boldsymbol{y}
-\mathbf{z}_s\,\hat{\mathbf b}_\xi
+\mathbf{z}_a\,\hat{\mathbf c}_\xi
-\boldsymbol{y}
\\
&=
(\mathbf{z}_a-\mathbf{z}_s)\hat{\mathbf p}_\xi
-2\boldsymbol{y}.
\end{align}
Similarly, for \(\hat{\mathbf q}_\xi\),
\begin{align}
\frac{d}{d\xi}\hat{\mathbf q}_\xi
&=
\frac{d}{d\xi}\hat{\mathbf b}_\xi
-
\frac{d}{d\xi}\hat{\mathbf c}_\xi
\\
&=
\mathbf{z}_a\,\hat{\mathbf b}_\xi
-\mathbf{z}_s\,\hat{\mathbf c}_\xi
-\boldsymbol{y}
+\mathbf{z}_s\,\hat{\mathbf b}_\xi
-\mathbf{z}_a\,\hat{\mathbf c}_\xi
+\boldsymbol{y}
\\
&=
(\mathbf{z}_a+\mathbf{z}_s)\hat{\mathbf q}_\xi.
\end{align}
The original coupled system is therefore decoupled into
\begin{equation}
\frac{d}{d\xi}\hat{\mathbf p}_\xi
=
(\mathbf{z}_a-\mathbf{z}_s)\hat{\mathbf p}_\xi
-2\boldsymbol{y},
\qquad
\frac{d}{d\xi}\hat{\mathbf q}_\xi
=
(\mathbf{z}_a+\mathbf{z}_s)\hat{\mathbf q}_\xi.
\end{equation}
Using the initial conditions $\hat{\mathbf b}_0=\hat{\mathbf a}$ and $\hat{\mathbf c}_0=\hat{\mathbf a}^\dagger$, we obtain
\begin{equation}
\hat{\mathbf p}_0=\hat{\mathbf a}+\hat{\mathbf a}^\dagger,
\qquad
\hat{\mathbf q}_0=\hat{\mathbf a}-\hat{\mathbf a}^\dagger.
\end{equation}
The equation for \(\hat{\mathbf q}_\xi\) is homogeneous with solution
\begin{equation}
\hat{\mathbf q}_\xi
=
e^{\xi(\mathbf{z}_a+\mathbf{z}_s)}(\hat{\mathbf a}-\hat{\mathbf a}^\dagger).
\end{equation}
The equation for \(\hat{\mathbf p}_\xi\) is inhomogeneous and its solution is
\begin{equation}
\hat{\mathbf p}_\xi
=
e^{\xi(\mathbf{z}_a-\mathbf{z}_s)}(\hat{\mathbf a}+\hat{\mathbf a}^\dagger)
-
2\int_0^\xi
e^{(\xi-t)(\mathbf{z}_a-\mathbf{z}_s)}
\boldsymbol{y}\,dt.
\end{equation}
Recovering $\hat{\mathbf b}_\xi$ and $\hat{\mathbf c}_\xi$ from
\begin{equation}
\hat{\mathbf b}_\xi=\frac12(\hat{\mathbf p}_\xi+\hat{\mathbf q}_\xi),
\qquad
\hat{\mathbf c}_\xi=\frac12(\hat{\mathbf p}_\xi-\hat{\mathbf q}_\xi),
\end{equation}
we obtain
\begin{equation}
\hat{\mathbf b}_\xi
=
\frac12\,e^{\xi(\mathbf{z}_a-\mathbf{z}_s)}(\hat{\mathbf a}+\hat{\mathbf a}^\dagger)
+
\frac12\,e^{\xi(\mathbf{z}_a+\mathbf{z}_s)}(\hat{\mathbf a}-\hat{\mathbf a}^\dagger)
-
\int_0^\xi
e^{(\xi-t)(\mathbf{z}_a-\mathbf{z}_s)}
\boldsymbol{y}\,dt,
\end{equation}
and
\begin{equation}
\hat{\mathbf c}_\xi
=
\frac12\,e^{\xi(\mathbf{z}_a-\mathbf{z}_s)}(\hat{\mathbf a}+\hat{\mathbf a}^\dagger)
-
\frac12\,e^{\xi(\mathbf{z}_a+\mathbf{z}_s)}(\hat{\mathbf a}-\hat{\mathbf a}^\dagger)
-
\int_0^\xi
e^{(\xi-t)(\mathbf{z}_a-\mathbf{z}_s)}
\boldsymbol{y}\,dt.
\end{equation}

Expanding the coefficients of \(\hat{\mathbf a}\) and \(\hat{\mathbf a}^\dagger\), we write the generalized Bogoliubov transformation for the transformed mode operators:
\begin{equation} \label{eq:b_xi_explicit_direct1}
\hat{\mathbf b}_\xi
=
\underbrace{\frac12\Bigl[
e^{\xi(\mathbf{z}_a-\mathbf{z}_s)}
+
e^{\xi(\mathbf{z}_a+\mathbf{z}_s)}
\Bigr]}_{\bar{\mathbf A}_\xi}\hat{\mathbf a}
+
\underbrace{\frac12\Bigl[
e^{\xi(\mathbf{z}_a-\mathbf{z}_s)}
-
e^{\xi(\mathbf{z}_a+\mathbf{z}_s)}
\Bigr]}_{\bar{\mathbf B}_\xi}\hat{\mathbf a}^\dagger
+
\underbrace{\int_0^\xi
-e^{(\xi-t)(\mathbf{z}_a-\mathbf{z}_s)}
\boldsymbol{y}\,dt}_{\bar{\mathbf u}_\xi},
\end{equation}
\begin{equation} \label{eq:b_xi_explicit_direct2}
\hat{\mathbf c}_\xi
=
\bar{\mathbf B}_\xi\hat{\mathbf a}
+
\bar{\mathbf A}_\xi\hat{\mathbf a}^\dagger
+
\bar{\mathbf u}_\xi.
\end{equation}

Using equations \eqref{eq:b_xi_explicit_direct1} and \eqref{eq:b_xi_explicit_direct2} we can show that the commutation relations become
\begin{align}
[\hat b_{\xi,i},\hat c_{\xi,j}]
&=
\left(
\bar{\mathbf A}_\xi\bar{\mathbf A}_\xi^T
-
\bar{\mathbf B}_\xi\bar{\mathbf B}_\xi^T
\right)_{ij}\hat I,
\\
[\hat b_{\xi,i},\hat b_{\xi,j}]
&=
\left(
\bar{\mathbf A}_\xi\bar{\mathbf B}_\xi^T
-
\bar{\mathbf B}_\xi\bar{\mathbf A}_\xi^T
\right)_{ij}\hat I,
\end{align}
where displacement terms $\bar{\mathbf u}_\xi$ does not contribute to these commutators. Therefore using equations \eqref{eq:comm_b_c} and
\eqref{eq:comm_b_b} we obtain
\begin{equation} \label{eq:commutation_raletions_symmetry}
\bar{\mathbf A}_\xi\bar{\mathbf A}_\xi^T
-
\bar{\mathbf B}_\xi\bar{\mathbf B}_\xi^T
=
\mathbf I,
\qquad
\bar{\mathbf A}_\xi\bar{\mathbf B}_\xi^T
=
\bar{\mathbf B}_\xi\bar{\mathbf A}_\xi^T.
\end{equation}

\subsection*{2d. Determining auxiliary operator parameters using the nullifier equation}

From equation \eqref{eq:b_xi_explicit_direct1} we obtain
\begin{equation}\label{eq:bog_aux_general}
e^{\xi\hat G_0}\hat{\mathbf a}\,e^{-\xi\hat G_0}
=
\bar{\mathbf A}_\xi \hat{\mathbf a}
+\bar{\mathbf B}_\xi \hat{\mathbf a}^\dagger
+\bar{\mathbf u}_\xi.
\end{equation}
Starting from the basic nullifier equation \(\hat{\mathbf a}|\mathbf 0\rangle=0\), one has
\begin{equation}
0
=
e^{\xi\hat G_0}\hat{\mathbf a}|\mathbf 0\rangle
=
\left(e^{\xi\hat G_0}\hat{\mathbf a}e^{-\xi\hat G_0}\right)e^{\xi\hat G_0}|\mathbf 0\rangle
=
\left(e^{\xi\hat G_0}\hat{\mathbf a}e^{-\xi\hat G_0}\right)|\psi_\xi^{(0)}\rangle.
\end{equation}
Using equation \eqref{eq:bog_aux_general}, this becomes
\begin{equation}
\left(
\bar{\mathbf A}_\xi \hat{\mathbf a}
+\bar{\mathbf B}_\xi \hat{\mathbf a}^\dagger
+\bar{\mathbf u}_\xi
\right)|\psi_\xi^{(0)}\rangle
=
0.
\end{equation}
We note that at $\xi=0$, one has $\bar{\mathbf A}_0=\mathbf I$, which is invertible. Since $\bar{\mathbf A}_\xi$ depends continuously on $\xi$, it remains invertible in a neighborhood of the origin, leading to the following nullifier equation
\begin{equation}\label{eq:annihilation_aux}
\left(
\hat{\mathbf a}
-\bar{\mathbf Z}_\xi \hat{\mathbf a}^\dagger
+\bar{\boldsymbol\eta}_\xi
\right)|\psi_\xi^{(0)}\rangle
=
0,
\end{equation}
where we introduced
\begin{equation}
\bar{\mathbf Z}_\xi=-\bar{\mathbf A}_\xi^{-1}\bar{\mathbf B}_\xi,
\qquad
\bar{\boldsymbol\eta}_\xi=\bar{\mathbf A}_\xi^{-1}\bar{\mathbf u}_\xi.
\end{equation}
It is clear that the nullifier equation \eqref{eq:annihilation_aux} is the one obtained previously in equation \eqref{eq:nullifier_squeezing_displaced}. This means that the matrix $\bar{\mathbf Z}_\xi$ determines the squeezing and rotation parameters of the auxiliary Gaussian state. Since $\bar{\mathbf A}_\xi$ is invertible, using equations in \eqref{eq:commutation_raletions_symmetry} one can show that $\bar{\mathbf Z}^T_\xi = -\bar{\mathbf B}^T_\xi\bar{\mathbf A}^{-T}_\xi = -\bar{\mathbf A}^{-1}_\xi\bar{\mathbf B}_\xi = \bar{\mathbf Z}_\xi$, meaning that $\bar{\mathbf Z}_\xi$ is a complex symmetric matrix. 

We restrict the auxiliary parameter $\xi$ to a neighborhood of the origin, in which the Gaussian evolution is well defined and the auxiliary state has finite and nonzero norm $||\bar{\mathbf Z}_\xi||^2 < 1$. Since displacement always preserves the norm of the state, to normalize the Gaussian state in equation \eqref{eq:annihilation_aux}, the matrix $\bar{\mathbf Z}_\xi$ is a positive definite matrix:
\begin{equation}\label{eq:aux_normalizability}
\mathbf I-\bar{\mathbf Z}_\xi
\bar{\mathbf Z}_\xi^\dagger>0.
\end{equation}
Hence, the Takagi singular values should satisfy
$0\leq\bar s_{j,\xi}<1$.
Since $\bar{\mathbf Z}_0=0$ and $\bar{\mathbf Z}_\xi$ is continuous near the origin, this strict inequality holds in a sufficiently small neighborhood of $\xi=0$. This means that $\bar{\mathbf Z}_\xi$ must admit the following Takagi factorization
\begin{equation}\label{eq:takagi_aux}
\bar{\mathbf Z}_\xi
=
\bar{\mathbf U}_\xi\,\tanh(\bar{\boldsymbol\Sigma}_\xi)\,\bar{\mathbf U}_\xi^T,
\end{equation}
where $\bar{\mathbf U}_\xi$ is unitary and
\begin{equation}
\bar{\boldsymbol\Sigma}_\xi
=
\mathrm{diag}(\bar{\sigma}_{1,\xi},\dots,\bar{\sigma}_{M,\xi}),
\qquad
\tanh(\bar{\sigma}_{j,\xi})=\bar{s}_{j,\xi},
\end{equation}
\(\bar{s}_{j,\xi}\) being the Takagi singular values of $\bar{\mathbf Z}_\xi$.
To eliminate the affine displacement term $\bar{\boldsymbol\eta}_\xi$, we assume that the state can be written as
\begin{equation}
|\psi_\xi^{(0)}\rangle=\hat D(\bar{\boldsymbol\gamma}_\xi)|\phi_\xi\rangle.
\end{equation}
Substitute this into equation \eqref{eq:annihilation_aux} and multiplying from the left by \(\hat D^\dagger(\bar{\boldsymbol\gamma}_\xi)\) one obtains
\begin{equation}
\hat D^\dagger(\bar{\boldsymbol\gamma}_\xi)
\left(
\hat{\mathbf a}
-\bar{\mathbf Z}_\xi\hat{\mathbf a}^\dagger
+\bar{\boldsymbol\eta}_\xi
\right)
\hat D(\bar{\boldsymbol\gamma}_\xi)|\phi_\xi\rangle=0.
\end{equation}
Using the displacement action \eqref{eq:Disp_action} we write
\begin{equation}
\hat D^\dagger(\bar{\boldsymbol\gamma}_\xi)\hat{\mathbf a}\hat D(\bar{\boldsymbol\gamma}_\xi)
=
\hat{\mathbf a}+\bar{\boldsymbol\gamma}_\xi,
\qquad
\hat D^\dagger(\bar{\boldsymbol\gamma}_\xi)\hat{\mathbf a}^\dagger\hat D(\bar{\boldsymbol\gamma}_\xi)
=
\hat{\mathbf a}^\dagger+\bar{\boldsymbol\gamma}_\xi^*,
\end{equation}
which leads to
\begin{equation}
\left[
\hat{\mathbf a}
-\bar{\mathbf Z}_\xi\hat{\mathbf a}^\dagger
+
\bigl(
\bar{\boldsymbol\gamma}_\xi-\bar{\mathbf Z}_\xi\bar{\boldsymbol\gamma}_\xi^*
+\bar{\boldsymbol\eta}_\xi
\bigr)
\right]|\phi_\xi\rangle=0.
\end{equation}
Therefore, choosing $\bar{\boldsymbol\gamma}_\xi=-\bar{\boldsymbol\eta}_\xi+\bar{\mathbf Z}_\xi\bar{\boldsymbol\gamma}_\xi^*$ eliminates the affine term and yields
\begin{equation}
\left(
\hat{\mathbf a}
-\bar{\mathbf Z}_\xi\hat{\mathbf a}^\dagger
\right)|\phi_\xi\rangle=0.
\end{equation}
Taking the complex conjugate of the equation for \(\bar{\boldsymbol\gamma}_\xi\)
\begin{equation}
\bar{\boldsymbol\gamma}_\xi^*
-\bar{\boldsymbol Z}_\xi^*\bar{\boldsymbol\gamma}_\xi
=
-\bar{\boldsymbol\eta}_\xi^*,
\end{equation}
one finds
\begin{equation}
\bar{\boldsymbol\gamma}_\xi
=
-(\mathbf I-\bar{\mathbf Z}_\xi\bar{\mathbf Z}_\xi^*)^{-1}
\left(
\bar{\boldsymbol\eta}_\xi+\bar{\mathbf Z}_\xi\bar{\boldsymbol\eta}_\xi^*
\right).
\end{equation}
Next, using the Takagi factorization in equation \eqref{eq:takagi_aux} and defining rotated mode operators
\begin{equation}\label{eq:rotated_mode_operators_c}
\hat{\mathbf d}
=
\bar{\mathbf U}_\xi^\dagger\hat{\mathbf a},
\qquad
\hat{\mathbf d}^\dagger
=
\bar{\mathbf U}_\xi^T\hat{\mathbf a}^\dagger,
\end{equation}
the nullifier equation becomes
\begin{equation}
\left(
\hat{\mathbf d}
-
\tanh(\bar{\boldsymbol\Sigma}_\xi)\hat{\mathbf d}^\dagger
\right)|\phi_\xi\rangle=0,
\end{equation}
or, mode by mode,
\begin{equation}
\left(
\hat d_j-\tanh(\bar{\sigma}_{j,\xi})\hat d_j^\dagger
\right)|\phi_\xi\rangle=0.
\end{equation}
This is precisely the nullifier equation of a product of single-mode squeezed vacuum states obtained in equation \eqref{eq:nullifier_squeezing3}. Therefore, according to our previously obtained result in \eqref{eq:multimode_squeezing_decomposition_2}, we can write the state $|\phi_\xi\rangle$ as the rotated squeezed vacuum state with a scalar prefactor
\begin{equation} \label{eq:unnormalized_state_phi}
|\phi_\xi\rangle
=
C_\xi\,\hat R(\bar{\mathbf U}_\xi)\hat S(\bar{\boldsymbol\Sigma}_\xi)|\mathbf 0\rangle,
\end{equation}
and the full state then writes
\begin{equation}
|\psi_\xi^{(0)}\rangle
=
C_\xi\,\hat D(\bar{\boldsymbol\gamma}_\xi)\hat R(\bar{\mathbf U}_\xi)\hat S(\bar{\boldsymbol\Sigma}_\xi)|\mathbf 0\rangle.
\end{equation}
This allows us to decompose the auxiliary Gaussian state \eqref{eq:auxiliary_Gaussian_state_general} into the building blocks of Gaussian boson sampling architecture
\begin{equation}\label{eq:aux_state_final}
e^{\xi\hat G}|\mathbf 0\rangle
=
e^{\xi\,\mathrm{Tr}(\mathbf{z})}\,C_\xi\,
\hat D(\bar{\boldsymbol\gamma}_\xi)\,
\hat R(\bar{\mathbf U}_\xi)\,
\hat S(\bar{\boldsymbol\Sigma}_\xi)\,
|\mathbf 0\rangle,
\end{equation}
where \(C_\xi\) is the auxiliary scalar prefactor.

\subsection*{2e. Determining the auxiliary scalar prefactor \(C_\xi\)}

The auxiliary scalar prefactor can be determined using vacuum overlaps as
\begin{equation} \label{eq:aux_scalar_prefactor}
C_\xi
=
\frac{\langle \mathbf 0| e^{\xi\hat G_0} |\mathbf 0\rangle}{\langle \mathbf 0|\hat D(\bar{\boldsymbol\gamma}_\xi)\hat R(\bar{\mathbf U}_\xi)\hat S(\bar{\boldsymbol\Sigma}_\xi)|\mathbf 0\rangle}
=
\frac{v_\xi}{g_\xi}.
\end{equation}
We first determine \(v_\xi\) from
\begin{equation}
\frac{d}{d\xi}|\psi_\xi^{(0)}\rangle
=
\hat G_0|\psi_\xi^{(0)}\rangle,
\end{equation}
which implies
\begin{equation}
\frac{d}{d\xi}v_\xi
=
\langle \mathbf 0|\hat G_0|\psi_\xi^{(0)}\rangle.
\end{equation}
Substituting \(\hat G_0\) and using
\begin{equation}
\langle \mathbf 0|\hat{\mathbf a}^\dagger=0,
\qquad
\langle \mathbf 0|\hat{\mathbf a}^{\dagger T}\mathbf z_s\hat{\mathbf a}^\dagger=0,
\qquad
\langle \mathbf 0|\hat{\mathbf a}^{\dagger T}\mathbf z_a\hat{\mathbf a}=0,
\end{equation}
we obtain
\begin{equation}\label{eq:c_xi_derivation_start}
\frac{d}{d\xi}v_\xi
=
-\boldsymbol y^T\langle \mathbf 0|\hat{\mathbf a}|\psi_\xi^{(0)}\rangle
-\frac12\langle \mathbf 0|\hat{\mathbf a}^T\mathbf z_s\hat{\mathbf a}|\psi_\xi^{(0)}\rangle.
\end{equation}

Using the nullifier equation \eqref{eq:annihilation_aux} and multiplying it from the left by \(\langle \mathbf 0|\), we can express the first term as
\begin{equation}
\langle \mathbf 0|\hat{\mathbf a}|\psi_\xi^{(0)}\rangle
=
-\bar{\boldsymbol\eta}_\xi\,v_\xi.
\end{equation}
To obtain the second term, we multiply the \(j\)-th component of equation \eqref{eq:annihilation_aux} from the left by \(\langle \mathbf 0|\hat a_i\), yielding
\begin{equation}
\langle \mathbf 0|\hat a_i\hat a_j|\psi_\xi^{(0)}\rangle
-
\sum_k (\bar Z_\xi)_{jk}\langle \mathbf 0|\hat a_i\hat a_k^\dagger|\psi_\xi^{(0)}\rangle
+
(\bar\eta_\xi)_j\langle \mathbf 0|\hat a_i|\psi_\xi^{(0)}\rangle
=
0.
\end{equation}
Using
\begin{equation}
\langle \mathbf 0|\hat a_i\hat a_k^\dagger
=
\delta_{ik}\langle \mathbf 0|,
\end{equation}
together with the first-moment relation, we obtain
\begin{equation}
\langle \mathbf 0|\hat a_i\hat a_j|\psi_\xi^{(0)}\rangle
=
\bigl[(\bar\eta_\xi)_i(\bar\eta_\xi)_j+(\bar Z_\xi)_{ji}\bigr]v_\xi.
\end{equation}
Since $\bar{\mathbf Z}_\xi$ is symmetric,
\begin{equation}
\langle \mathbf 0|\hat{\mathbf a}\hat{\mathbf a}^T|\psi_\xi^{(0)}\rangle
=
\left(
\bar{\boldsymbol\eta}_\xi\bar{\boldsymbol\eta}_\xi^T+\bar{\mathbf Z}_\xi
\right)v_\xi.
\end{equation}
Substituting these expressions into equation \eqref{eq:c_xi_derivation_start}, we obtain
\begin{equation}
\frac{d}{d\xi}v_\xi
=
\left[
\boldsymbol y^T\bar{\boldsymbol\eta}_\xi
-\frac12\,\mathrm{Tr}(\mathbf z_s\bar{\mathbf Z}_\xi)
-\frac12\,\bar{\boldsymbol\eta}_\xi^T\mathbf z_s\bar{\boldsymbol\eta}_\xi
\right]v_\xi.
\end{equation}
With the initial condition \(v_0=1\), the solution is
\begin{equation}\label{eq:c_xi_final}
v_\xi
=
\exp\!\left[
\int_0^\xi
\left(
\boldsymbol y^T\bar{\boldsymbol\eta}_t
-\frac12\,\mathrm{Tr}(\mathbf z_s\bar{\mathbf Z}_t)
-\frac12\,\bar{\boldsymbol\eta}_t^T\mathbf z_s\bar{\boldsymbol\eta}_t
\right)\,dt
\right].
\end{equation}

To evaluate $g_\xi$, we start from
\begin{equation}
g_\xi
=
\langle \mathbf 0|
\hat D(\bar{\boldsymbol\gamma}_\xi)\hat R(\bar{\mathbf U}_\xi)\hat S(\bar{\boldsymbol\Sigma}_\xi)
|\mathbf 0\rangle,
\end{equation}
and then we swap the displacement and rotation operators
\begin{equation}
\hat D(\bar{\boldsymbol\gamma}_\xi)\hat R(\bar{\mathbf U}_\xi)
=
\hat R(\bar{\mathbf U}_\xi)\hat D(\bar{\mathbf U}^\dagger_\xi\bar{\boldsymbol\gamma}_\xi),
\end{equation}
together with the invariance of the vacuum under passive rotations, we obtain
\begin{equation}
g_\xi
=
\langle \mathbf 0|
\hat D(\bar{\boldsymbol\gamma}^{(r)}_\xi)\hat S(\bar{\boldsymbol\Sigma}_\xi)
|\mathbf 0\rangle,
\qquad
\bar{\boldsymbol\gamma}^{(r)}_\xi=\bar{\mathbf U}^\dagger_\xi\bar{\boldsymbol\gamma}_\xi.
\end{equation}

Using the Takagi factorization of $\bar{\mathbf Z}_\xi$,
\begin{equation}
\bar{\mathbf Z}_\xi
=
\bar{\mathbf U}_\xi\,\mathrm{diag}(\bar s_{1,\xi},\dots,\bar s_{M,\xi})\,\bar{\mathbf U}_\xi^T,
\qquad
0\le \bar s_{j,\xi}<1,
\end{equation}
together with the rotated mode operators in equation \eqref{eq:rotated_mode_operators_c}, we obtain
\begin{equation}
\hat{\mathbf a}^{\dagger T}\bar{\mathbf Z}_\xi\hat{\mathbf a}^\dagger
=
\hat{\mathbf d}^{\dagger T}\mathrm{diag}(\bar s_{1,\xi},\dots,\bar s_{M,\xi})\hat{\mathbf d}^\dagger
=
\sum_{j=1}^M \bar s_{j,\xi}\hat d_j^{\dagger 2}.
\end{equation}
Hence, we can define a normalized Gaussian state $|\chi_\xi\rangle$ that represents only the multimode squeezed part $|\chi_\xi\rangle
= \hat R(\bar{\mathbf U}_\xi) \hat S(\bar{\boldsymbol\Sigma}_\xi)|\mathbf 0\rangle$, which is related to the equation \eqref{eq:unnormalized_state_phi} as $|\phi_\xi\rangle = C_\xi|\chi_\xi\rangle$. This state can be expressed as
\begin{equation}
|\chi_\xi\rangle
=
N_\xi\exp\!\left(\frac12\hat{\mathbf a}^{\dagger T}\bar{\mathbf Z}_\xi\hat{\mathbf a}^\dagger\right)|\mathbf 0\rangle
\end{equation}
where $N_\xi$ is the normalization factor that needs to be determined. We already know that this state can be factorized into a product of single-mode states in the rotated mode basis using equation \eqref{eq:rotated_mode_operators_c} as
\begin{equation}
|\chi_\xi\rangle
=
N_\xi\prod_{j=1}^M
\exp\!\left(\frac12 \bar s_{j,\xi}\hat d_j^{\dagger 2}\right)|\mathbf 0\rangle.
\end{equation}
We now normalize each single-mode factor separately. Starting from 
\begin{equation}
|\chi_s\rangle
=
N_s\exp\!\left(\frac12 s\,\hat d^{\dagger 2}\right)|0\rangle,
\end{equation}
and expanding the exponential gives
\begin{equation}
\exp\!\left(\frac12 s\,\hat d^{\dagger 2}\right)|0\rangle
=
\sum_{n=0}^\infty
\frac{1}{n!}
\left(\frac{s}{2}\right)^n
\hat d^{\dagger 2n}|0\rangle.
\end{equation}
Using
\begin{equation}
\hat d^{\dagger 2n}|0\rangle
=
\sqrt{(2n)!}\,|2n\rangle,
\end{equation}
we obtain
\begin{equation}
|\chi_s\rangle
=
N_s
\sum_{n=0}^\infty
\frac{1}{n!}
\left(\frac{s}{2}\right)^n
\sqrt{(2n)!}\,|2n\rangle.
\end{equation}
The norm is therefore
\begin{equation}
\langle \chi_s|\chi_s\rangle
=
|N_s|^2
\sum_{n=0}^\infty
\frac{(2n)!}{(n!)^2\,4^n}|s|^{2n}.
\end{equation}
Using
\begin{equation}
\sum_{n=0}^\infty
\frac{(2n)!}{(n!)^2\,4^n}x^n
=
\frac{1}{\sqrt{1-x}},
\end{equation}
we find
\begin{equation}
\langle \chi_s|\chi_s\rangle
=
|N_s|^2(1-|s|^2)^{-1/2}.
\end{equation}
Imposing normalization, \(\langle \chi_s|\chi_s\rangle=1\), yields
\begin{equation}
N_s=(1-|s|^2)^{1/4},
\end{equation}
up to an irrelevant phase. For the multimode state, the normalization constant is therefore
\begin{equation}
N_\xi
=
\prod_{j=1}^M (1-\bar s_{j,\xi}^2)^{1/4}.
\end{equation}
Returning to the expression for $g_\xi$ with the rotated
displacement, we use
\begin{equation}
\hat S(\bar{\boldsymbol\Sigma}_\xi)|\mathbf 0\rangle
=
\hat R^\dagger(\bar{\mathbf U}_\xi)|\chi_\xi\rangle
=
N_\xi
\exp\!\left(
\frac12\hat{\mathbf a}^{\dagger T}
\bar{\mathbf S}_\xi
\hat{\mathbf a}^\dagger
\right)|\mathbf 0\rangle,
\qquad
\bar{\mathbf S}_\xi
=
\mathrm{diag}(\bar s_{1,\xi},\dots,\bar s_{M,\xi}).
\end{equation}
Therefore,
\begin{equation}
g_\xi
=
\det(\mathbf{I} -\bar{\mathbf{S}}_{\xi}^2)^{1/4}
\left\langle \mathbf 0\left|
\hat D(\bar{\boldsymbol\gamma}^{(r)}_\xi)
\exp\!\left(\frac12\hat{\mathbf a}^{\dagger T}\bar{\mathbf{S}}_{\xi}\hat{\mathbf a}^\dagger\right)
\right|\mathbf 0\right\rangle.
\end{equation}
Using the Baker–Campbell–Hausdorff decomposition of the displacement operator
\begin{equation}
\hat D(\bar{\boldsymbol\gamma}^{(r)}_\xi)
=
\exp\!\left(-\frac12\bar{\boldsymbol\gamma}^{(r)\dagger}_\xi\bar{\boldsymbol\gamma}^{(r)}_\xi\right)
\exp\!\left(\bar{\boldsymbol\gamma}^{(r)T}_\xi\hat{\mathbf a}^\dagger\right)
\exp\!\left(-\bar{\boldsymbol\gamma}^{(r)\dagger}_\xi\hat{\mathbf a}\right),
\end{equation}
and using \(\langle \mathbf 0|\hat{\mathbf a}^\dagger=0\), we find
\begin{equation}
g_\xi
=
\det(\mathbf{I} -\bar{\mathbf{S}}_{\xi}^2)^{1/4}
\exp\!\left(-\frac12\bar{\boldsymbol\gamma}^{(r)\dagger}_\xi\bar{\boldsymbol\gamma}^{(r)}_\xi\right)
\left \langle \mathbf 0 \left|
\exp\!\left(-\bar{\boldsymbol\gamma}^{(r)\dagger}_\xi\hat{\mathbf a}\right)
\exp\!\left(\frac12\hat{\mathbf a}^{\dagger T}\bar{\mathbf S}_\xi\hat{\mathbf a}^\dagger\right)
\right|\mathbf 0 \right\rangle .
\end{equation}
To evaluate the remaining vacuum expectation value we define
\begin{equation}
\hat K
=
-\bar{\boldsymbol\gamma}^{(r)\dagger}_\xi\hat{\mathbf a},
\qquad
\hat L
=
\frac12
\hat{\mathbf a}^{\dagger T}
\bar{\mathbf S}_\xi
\hat{\mathbf a}^\dagger .
\end{equation}
Since \(\bar{\mathbf S}_\xi\) is symmetric, one finds
\begin{equation}
[\hat K,\hat L]
=
-\bar{\boldsymbol\gamma}^{(r)\dagger}_\xi
\bar{\mathbf S}_\xi
\hat{\mathbf a}^\dagger ,
\qquad
[\hat K,[\hat K,\hat L]]
=
\bar{\boldsymbol\gamma}^{(r)\dagger}_\xi
\bar{\mathbf S}_\xi
\bar{\boldsymbol\gamma}^{(r)*}_\xi .
\end{equation}
Since \(\hat L\) and $[\hat K,\hat L]$ contain only creation operators and therefore commute, we write
\begin{equation}
e^{\hat K}e^{\hat L}
=
\exp(\hat L)
\exp([\hat K,\hat L])
\exp\!\left(
\frac12[\hat K,[\hat K,\hat L]]
\right)
\exp(\hat K),
\end{equation}
Next, we use
\begin{equation}
\langle \mathbf 0|e^{\hat L}=\langle \mathbf 0|,
\qquad
e^{\hat K}|\mathbf 0\rangle=|\mathbf 0\rangle,
\qquad
\langle \mathbf 0|
\exp([\hat K,\hat L])
|\mathbf 0\rangle
=
1,
\end{equation}
and obtain
\begin{equation}
\left\langle \mathbf 0\left|
\exp\!\left(
-\bar{\boldsymbol\gamma}^{(r)\dagger}_\xi\hat{\mathbf a}
\right)
\exp\!\left(
\frac12
\hat{\mathbf a}^{\dagger T}
\bar{\mathbf S}_\xi
\hat{\mathbf a}^\dagger
\right)
\right|\mathbf 0\right\rangle
=
\exp\!\left[
\frac12
\bar{\boldsymbol\gamma}^{(r)\dagger}_\xi
\bar{\mathbf S}_\xi
\bar{\boldsymbol\gamma}^{(r)*}_\xi
\right].
\end{equation}
Hence, the final form of $g_\xi$ writes
\begin{equation}\label{eq:g_xi_final}
g_\xi
=
\det(\mathbf{I}-\bar{\mathbf{S}}_{\xi}^2)^{1/4}
\exp\!\left[
\frac12\bar{\boldsymbol\gamma}^{(r)\dagger}_\xi\bar{\mathbf{S}}_\xi\bar{\boldsymbol\gamma}^{(r)*}_\xi 
-
\frac12\bar{\boldsymbol\gamma}^{(r)\dagger}_\xi\bar{\boldsymbol\gamma}^{(r)}_\xi
\right].
\end{equation}
Notably, an equivalent form of $g_\xi$ can be written in the original basis with no rotation as
\begin{equation}\label{eq:g_xi_final_unrotated}
g_\xi
=
\det(\mathbf{I} -\bar{\mathbf Z}_\xi\bar{\mathbf Z}_\xi^*)^{1/4}
\exp\!\left[
\frac12\bar{\boldsymbol\gamma}^\dagger_\xi\bar{\mathbf Z}_\xi\bar{\boldsymbol\gamma}^*_\xi 
-
\frac12\bar{\boldsymbol\gamma}^\dagger_\xi\bar{\boldsymbol\gamma}_\xi
\right].
\end{equation}
which is the form that we use in the main article.

\section*{3. Merging the auxiliary operator with Doktorov operator}

At this stage, the auxiliary displacement, rotation and squeezing operators should be combined with the ones from the Doktorov part. We start from
\begin{equation} \label{eq:Aux_f3}
f_{m}(\xi)
=
|\langle \boldsymbol m|\hat{U}_{Dok} e^{\xi\hat G}|\mathbf 0\rangle|^2,
\end{equation}
and using equation \eqref{eq:aux_state_final} with expanded \(\hat U_{Dok}\) one gets
\begin{equation} \label{eq:Aux_f4}
f_{m}(\xi)
=
|e^{\xi\,\mathrm{Tr}(\mathbf{z})}|^2|C_\xi|^2
\left|
\langle \boldsymbol m|
\hat D(\boldsymbol\beta)\hat R(\mathbf{U_L})\hat S(\boldsymbol r)\hat R(\mathbf{U_R})
\hat D(\bar{\boldsymbol\gamma}_\xi)\hat R(\bar{\mathbf U}_\xi)\hat S(\bar{\boldsymbol\Sigma}_\xi)
|\mathbf 0\rangle
\right|^2.
\end{equation}
To simplify the notation, we define the Gaussian part of the Doktorov operator as
\begin{equation}\label{eq:U0_def_merge}
\hat U_0
=
\hat R(\mathbf{U_L})\hat S(\boldsymbol r)\hat R(\mathbf{U_R}),
\qquad
\hat U_{Dok}=\hat D(\boldsymbol\beta)\hat U_0.
\end{equation}
leading to this form of the auxiliary function
\begin{equation}\label{eq:Aux_f5}
f_{m}(\xi)
=
|e^{\xi\,\mathrm{Tr}(\mathbf{z})}|^2|C_\xi|^2
\left|
\left\langle \boldsymbol{m} \left|
\hat D(\boldsymbol\beta)\,
\hat U_0\,
\hat D(\bar{\boldsymbol\gamma}_\xi)\,
\hat R(\bar{\mathbf U}_\xi)\,
\hat S(\bar{\boldsymbol\Sigma}_\xi)\,
\right|\mathbf 0\right\rangle
\right|^2.
\end{equation}
The auxiliary state under consideration
\begin{equation}
|\psi_\xi^{(0)}\rangle
=
C_\xi
\hat D(\bar{\boldsymbol\gamma}_\xi)\,
\hat R(\bar{\mathbf U}_\xi)\,
\hat S(\bar{\boldsymbol\Sigma}_\xi)\,
|\mathbf 0\rangle
\end{equation}
satisfies the nullifier equation
\begin{equation}\label{eq:annihilation_aux_repeat}
\left(
\hat{\mathbf a}
-\bar{\mathbf Z}_\xi \hat{\mathbf a}^\dagger
+\bar{\boldsymbol\eta}_\xi
\right)|\psi_\xi^{(0)}\rangle
=
0.
\end{equation}
Thus, one can write the transformed auxiliary state as
\begin{equation}\label{eq:psi_tilde_def}
|\tilde \psi_\xi\rangle
=
\hat U_0 |\psi_\xi^{(0)}\rangle.
\end{equation}
To determine its operator parameters, one needs to find the parameters in the following nullifier equation
\begin{equation}\label{eq:annihilation_tilde}
\left(
\hat{\mathbf a}
-\mathbf Z_\xi \hat{\mathbf a}^\dagger
+\boldsymbol\eta_\xi
\right)
|\tilde \psi_\xi\rangle
=
0.
\end{equation}
Note that we use letters without bar signs for the new nullifier equation. Using \eqref{eq:psi_tilde_def} this equation can be conjugated by \(\hat U_0^\dagger\) to obtain
\begin{equation}
\hat U_0^\dagger
\left(
\hat{\mathbf a}
-\mathbf Z_\xi \hat{\mathbf a}^\dagger
+\boldsymbol\eta_\xi
\right)
\hat U_0
|\psi_\xi^{(0)}\rangle
=
0.
\end{equation}
Transformation of the ladder operators by $\hat U_0$ can be written in Bogoliubov form
\begin{equation}\label{eq:Bogoliubov_U0_merge}
\hat U_0^\dagger\,\hat{\mathbf a}\,\hat U_0
=
\mathbf X\,\hat{\mathbf a}+\mathbf Y\,\hat{\mathbf a}^\dagger,
\qquad
\hat U_0^\dagger\,\hat{\mathbf a}^\dagger\,\hat U_0
=
\mathbf Y^*\,\hat{\mathbf a}+\mathbf X^*\,\hat{\mathbf a}^\dagger,
\end{equation}
with
\begin{equation}\label{eq:XY_merge}
\mathbf X=\mathbf{U_L}\,\cosh(\boldsymbol r)\,\mathbf{U_R},
\qquad
\mathbf Y=\mathbf{U_L}\,\sinh(\boldsymbol r)\,\mathbf{U_R}^*,
\end{equation}
hence, we obtain
\begin{equation}
\left[
\left(\mathbf X-\mathbf Z_\xi\mathbf Y^*\right)\hat{\mathbf a}
+
\left(\mathbf Y-\mathbf Z_\xi\mathbf X^*\right)\hat{\mathbf a}^\dagger
+
\boldsymbol\eta_\xi
\right]
|\psi_\xi^{(0)}\rangle
=
0.
\end{equation}
Comparing this equation with the auxiliary nullifier \eqref{eq:annihilation_aux_repeat}, one finds
\begin{equation}
\mathbf Y-\mathbf Z_\xi\mathbf X^*
=
-\left(\mathbf X-\mathbf Z_\xi\mathbf Y^*\right)\bar{\mathbf Z}_\xi,
\qquad
\boldsymbol\eta_\xi
=
\left(\mathbf X-\mathbf Z_\xi\mathbf Y^*\right)\bar{\boldsymbol\eta}_\xi.
\end{equation}
Rearranging the first relation gives
\begin{equation}
\mathbf Z_\xi
\left(
\mathbf X^*+\mathbf Y^*\bar{\mathbf Z}_\xi
\right)
=
\mathbf X\bar{\mathbf Z}_\xi+\mathbf Y.
\end{equation}
We now introduce
\begin{equation}\label{eq:Atilde_Btilde_def}
\mathbf A_\xi
=
\mathbf X^*+\mathbf Y^*\bar{\mathbf Z}_\xi,
\qquad
\mathbf B_\xi
=
\mathbf Y+\mathbf X\bar{\mathbf Z}_\xi.
\end{equation}
Since our auxiliary Gaussian state is normalizable, it implies $\mathbf I-\bar{\mathbf Z}_\xi^\dagger\bar{\mathbf Z}_\xi>0$. Together with the Bogoliubov identities, this gives $\mathbf A_\xi^\dagger\mathbf A_\xi=\mathbf B_\xi^\dagger\mathbf B_\xi+\mathbf I-\bar{\mathbf Z}_\xi^\dagger\bar{\mathbf Z}_\xi>0$, meaning that $\mathbf A_\xi$ is invertible, thus one can write
\begin{equation}\label{eq:Zeta_tilde_def}
\mathbf Z_\xi
=
\mathbf B_\xi\mathbf A_\xi^{-1},
\qquad
\boldsymbol\eta_\xi
=
\left(\mathbf X-\mathbf Z_\xi\mathbf Y^*\right)\bar{\boldsymbol\eta}_\xi.
\end{equation}
Equation \eqref{eq:annihilation_tilde} shows that the transformed auxiliary state remains a pure displaced Gaussian state. Since $\hat U_0$ is unitary, its norm is
unchanged, and the state can therefore be written as
\begin{equation}\label{eq:psi_tilde_final_form}
|\widetilde\psi_\xi\rangle
=
C_\xi e^{i\vartheta_\xi}
\hat D(\boldsymbol\gamma_\xi)
\hat R(\mathbf U_\xi)
\hat S(\boldsymbol\Sigma_\xi)
|\mathbf0\rangle,
\end{equation}
where $\vartheta_\xi\in\mathbb R$ is an overall phase.

The Bogoliubov identities and the symmetry of
$\bar{\mathbf Z}_\xi$ imply
$\mathbf A_\xi^T\mathbf B_\xi
=\mathbf B_\xi^T\mathbf A_\xi$.
Since $\mathbf A_\xi$ is invertible, it follows that
\begin{equation}
\mathbf Z_\xi^T
=
\mathbf A_\xi^{-T}\mathbf B_\xi^T
=
\mathbf B_\xi\mathbf A_\xi^{-1}
=
\mathbf Z_\xi.
\end{equation}
Thus, $\mathbf Z_\xi$ is complex symmetric and must admit
the Takagi factorization. More importantly, this factorization directly provides the operator parameters for the rotation and squeezing operators in equation \eqref{eq:psi_tilde_final_form}:
\begin{equation}\label{eq:takagi_tilde}
\mathbf Z_\xi
=
\mathbf U_\xi\,
\tanh(\boldsymbol\Sigma_\xi)\,
\mathbf U_\xi^T.
\end{equation}
The effective displacement is
\begin{equation}\label{eq:gamma_tilde_def}
\boldsymbol\gamma_\xi
=
-(\mathbf I-\mathbf Z_\xi\mathbf Z_\xi^*)^{-1}
\left(
\boldsymbol\eta_\xi
+
\mathbf Z_\xi\boldsymbol\eta_\xi^*
\right).
\end{equation}
Substituting equation \eqref{eq:psi_tilde_final_form} into equation \eqref{eq:Aux_f5}, we obtain
\begin{equation}\label{eq:Aux_f6}
f_{m}(\xi)
=
|e^{\xi\,\mathrm{Tr}(\mathbf{z})}|^2|C_\xi|^2
\left|
\left\langle \boldsymbol{m}\left|
\hat D(\boldsymbol\beta)\,
\hat D(\boldsymbol\gamma_\xi)\,
\hat R(\mathbf U_\xi)\,
\hat S(\boldsymbol\Sigma_\xi)\,
\right|\mathbf 0\right\rangle
\right|^2,
\end{equation}
where the overall phase $e^{i\vartheta_\xi}$ in \eqref{eq:psi_tilde_final_form} disappears upon taking the modulus squared.

Using the Weyl composition rule
\begin{equation}
\hat D(\boldsymbol\beta)\hat D(\boldsymbol\gamma_\xi)
=
e^{-i\,\mathrm{Im}(\boldsymbol\beta^\dagger\boldsymbol\gamma_\xi)}
\hat D(\boldsymbol\beta+\boldsymbol\gamma_\xi),
\end{equation}
and noting that the phase disappears after taking the modulus squared, we find the total effective displacement
\begin{equation}\label{eq:beta_eff_merge}
\boldsymbol\beta_\xi
=
\boldsymbol\beta+\boldsymbol\gamma_\xi,
\end{equation}
so that the final form of the auxiliary function is
\begin{equation}\label{eq:Aux_f_final_merge}
f_{m}(\xi)
=
|e^{\xi\,\mathrm{Tr}(\mathbf{z})}|^2|C_\xi|^2
\left|
\left\langle \boldsymbol{m}\left|
\hat D(\boldsymbol\beta_\xi)\,
\hat R(\mathbf U_\xi)\,
\hat S(\boldsymbol\Sigma_\xi)\,
\right|\mathbf 0\right\rangle
\right|^2.
\end{equation}

\section*{4. Condon approximation for NAC matrix element calculation}

In the Condon approximation, the auxiliary operator contains only the linear term,
\begin{equation}
\hat G
=
\sum_{k=1}^M y_k(\hat a_k^\dagger-\hat a_k),
\end{equation}
so that the quadratic contribution vanishes:
\begin{equation}
\boldsymbol z=0,
\qquad
\mathrm{Tr}(\mathbf z)=0,
\qquad
\boldsymbol z_s=0,
\qquad
\boldsymbol z_a=0.
\end{equation}
The auxiliary state then reads
\begin{equation}
e^{\xi \hat G}|\mathbf 0\rangle
=
C_\xi\,\hat D(\bar{\boldsymbol\gamma}_\xi)|\mathbf 0\rangle,
\end{equation}
with no auxiliary squeezing and rotation. Hence,
\begin{equation}
\bar{\mathbf Z}_\xi=0,
\qquad
\bar{\boldsymbol\Sigma}_\xi=0,
\qquad
\bar{\mathbf U}_\xi=\mathbf I,
\qquad
\bar{\boldsymbol\gamma}_\xi=\xi\boldsymbol y.
\end{equation}
Therefore,
\begin{equation}
e^{\xi\hat G}|\mathbf 0\rangle
=
C_\xi\,\hat D(\xi\boldsymbol y)|\mathbf 0\rangle.
\end{equation}
Since \(\bar{\mathbf Z}_\xi=0\), equations \eqref{eq:Atilde_Btilde_def} reduce to
\begin{equation}
\mathbf A_\xi=\mathbf X^*,
\qquad
\mathbf B_\xi=\mathbf Y,
\end{equation}
and therefore
\begin{equation}
\mathbf Z_\xi
=
\mathbf B_\xi \mathbf A_\xi^{-1}
=
\mathbf Y(\mathbf X^*)^{-1}.
\end{equation}
This matrix is independent of $\xi$ and thus is associated with the Doktorov Gaussian unitary $\hat U_0$. In other words, no new auxiliary squeezing structure is generated in the Condon case.

Likewise we have
\begin{equation}
\boldsymbol\eta_\xi
=
\left(\mathbf X-\mathbf Z_\xi\mathbf Y^*\right)\bar{\boldsymbol\eta}_\xi.
\end{equation}
Since in the Condon case the auxiliary state is purely displaced, one may equivalently commute the displacement operator directly through \(\hat U_0\):
\begin{equation}
\hat U_0\hat D(\xi\boldsymbol y)
=
e^{i\phi(\xi\boldsymbol y)}
\hat D(\boldsymbol\eta')\hat U_0,
\qquad
\boldsymbol\eta'=\mathbf X(\xi\boldsymbol y)+\mathbf Y(\xi\boldsymbol y)^*.
\end{equation}
After merging this displacement with the Doktorov displacement \(\hat D(\boldsymbol\beta)\), one obtains
\begin{equation}
\boldsymbol\beta_\xi
=
\boldsymbol\beta+\boldsymbol\eta'
=
\boldsymbol\beta+\mathbf X(\xi\boldsymbol y)+\mathbf Y(\xi\boldsymbol y)^*.
\end{equation}
Therefore, in the Condon approximation, equation \eqref{eq:Aux_f_final_merge} reduces to
\begin{equation}
f_m(\xi)
=
|C_\xi|^2
\left|
\left\langle \boldsymbol{m}\left|
\hat D(\boldsymbol\beta_\xi)\,
\hat R(\mathbf{U_L})\hat S(\boldsymbol r)
\right|\mathbf 0\right\rangle
\right|^2,
\end{equation}
where we used \(\mathrm{Tr}(\boldsymbol z)=0\) and the fact that the rightmost rotation acts trivially on the vacuum.

In the Condon approximation, the auxiliary prefactor \(C_\xi\) simplifies considerably because the quadratic auxiliary contribution vanishes:
\begin{equation}
\mathbf z=0,
\qquad
\mathbf z_s=0,
\qquad
\bar{\mathbf{Z}}_\xi=0.
\end{equation}
The general expression for \(v_\xi\) then reduces to
\begin{equation}
v_\xi
=
\exp\!\left[
\int_0^\xi
\boldsymbol y^T\bar{\boldsymbol\eta}_t\,dt
\right].
\end{equation}
In the Condon case,
\begin{equation}
e^{\xi\hat G}|\mathbf 0\rangle
=
C_\xi\,\hat D(\xi\boldsymbol y)|\mathbf 0\rangle,
\end{equation}
with \(\bar{\boldsymbol\eta}_\xi=-\xi\boldsymbol y\), hence \(\bar{\boldsymbol\eta}_t=-t\boldsymbol y\). Therefore,
\begin{align}
v_\xi
&=
\exp\!\left[
\int_0^\xi
-t\,\boldsymbol y^T\boldsymbol y\,dt
\right]
\\
&=
\exp\!\left(
-\frac{\xi^2}{2}\,\boldsymbol y^T\boldsymbol y
\right).
\end{align}

For the denominator \(g_\xi\), equation \eqref{eq:g_xi_final} simplifies because \(\bar{\mathbf{Z}}_\xi=0\) and \(\bar{\boldsymbol\gamma}_\xi=\xi\boldsymbol y\), yielding
\begin{equation}
g_\xi
=
\exp\!\left(
-\frac12(\xi\boldsymbol y)^\dagger(\xi\boldsymbol y)
\right).
\end{equation}
Hence,
\begin{equation}
C_\xi=\frac{v_\xi}{g_\xi}
=
\exp\!\left[
\frac12\left(
(\xi\boldsymbol y)^\dagger(\xi\boldsymbol y)
-
\xi^2\,\boldsymbol y^T\boldsymbol y
\right)
\right].
\end{equation}
For real \(\boldsymbol y\), this becomes
\begin{equation}
C_\xi
=
\exp\!\left[
\frac12\left(
|\xi|^2-\xi^2
\right)\boldsymbol y^T\boldsymbol y
\right].
\end{equation}

For the finite-difference points
\begin{equation}
\xi \in \{0,\tau,-\tau,i\tau\},
\qquad \tau\in\mathbb R,
\end{equation}
one finds
\begin{equation}
C_{0}=1,
\qquad
C_{\tau}=1,
\qquad
C_{-\tau}=1,
\end{equation}
while for \(\xi=i\tau\),
\begin{equation}
C_{i\tau}
=
\exp\!\left(\tau^2\,\boldsymbol y^T\boldsymbol y\right).
\end{equation}
Thus, within the Condon approximation, the auxiliary prefactor contributes only for the imaginary displacement point \(\xi=i\tau\), whereas for the real points \(\xi=0,\pm\tau\) it reduces to unity.

\bibliographystyle{naturemag.bst}
\bibliography{References}